\documentclass[12pt]{article}
\usepackage{graphicx}
\usepackage{psfrag}
\usepackage{amsmath}
\usepackage{amsfonts}
\usepackage{amssymb}

\def\beq{\begin{equation}}
\def\eq{\end{equation}}
\def\eeq{\end{equation}}

\def\centeron#1#2{{\setbox0=\hbox{#1}\setbox1=\hbox{#2}\ifdim
\wd1>\wd0\kern.5\wd1\kern-.5\wd0\fi
\copy0\kern-.5\wd0\kern-.5\wd1\copy1\ifdim\wd0>\wd1
\kern.5\wd0\kern-.5\wd1\fi}}
\def\ltap{\;\centeron{\raise.35ex\hbox{$<$}}{\lower.65ex\hbox{$\sim$}}\;}
\def\gtap{\;\centeron{\raise.35ex\hbox{$>$}}{\lower.65ex\hbox{$\sim$}}\;}
\def\gsim{\mathrel{\gtap}}
\def\lsim{\mathrel{\ltap}}

\def\chii0{\chi_i^0}
\def\chij0{\chi_j^0}

\newcommand{\nc}{\newcommand}
\nc{\barrayn}{\begin{eqnarray*}}
\nc{\earrayn}{\end{eqnarray*}}
\nc{\er}[1]{(\ref{eq:#1})}
\nc{\ttbar}{t\bar  t}
\nc{\spacer}{\phantom{spacer}}
\nc{\ev}{\;\mathrm{eV}}
\nc{\mev}{\;\mathrm{MeV}}
\nc{\gev}{\;\mathrm{GeV}}
\nc{\tev}{\;\mathrm{TeV}}
\nc{\mc}{\mathcal}
\nc{\mcP}{\mc{P}}
\nc{\hc}{\;\mathrm{H.c.}}
\nc{\mcL}{\mathcal{L}}
\nc{\mcM}{\mathcal{M}}
\nc{\Z}{{\mathbb Z }}

\def\sun{\odot}

\def\PFGstripminus-#1{#1}%
\def\PFGshift(#1,#2)#3{\raisebox{#2}[\height][\depth]{\hbox{%
  \ifdim#1<0pt\kern#1 #3\kern\PFGstripminus#1\else\kern#1 #3\kern-#1\fi}}}%
\newcommand{\fref}[1]{Fig.\ \ref{f.#1}} 
\newcommand{\eref}[1]{Eq.\ (\ref{e.#1})}

\newcommand{\sref}[1]{Section \ref{s.#1}}
\newcommand{\cref}[1]{Chapter \ref{c.#1}}

\def\beq{\begin{equation}} 
\def\eeq{\end{equation}} 
\newcommand{\ba}{\begin{array}}  
\newcommand{\ea}{\end{array}} 
\newcommand{\bea}{\begin{eqnarray}}  
\newcommand{\eea}{\end{eqnarray} }  
\newcommand{\bal}{\begin{align}}
\newcommand{\eal}{\end{align}}   
\def\bi{\begin{itemize}}  
\def\ei{\end{itemize}}  
\def\ben{\begin{enumerate}}  
\def\een{\end{enumerate}}  
\def\beq{\begin{equation}}  
\def\eeq{\end{equation}}  
\def\bc{\begin{center}}
\def\ec{\end{center}} 
 \def\bt{\begin{table}}
\def\et{\end{table}}  
 \def\btb{\begin{tabular}}
\def\etb{\end{tabular}}

\def\cl{{\mathcal L}}

\def\ev{\, {\rm  eV}}
\def\gev{\, {\rm GeV}}
\def\tev{\, {\rm TeV}}

\def\mass2{mass${}^2$}

\def\ra{\rangle}
\def\la{\langle}  

\def\pa{\partial}

\newcommand{\tr}{\mathrm T \mathrm r}

\newcommand\simlt{\stackrel{<}{{}_\sim}}

\def\hc{{\rm h.c.}}

\def\ov{\overline}

\newcommand{\cm}{\mathrm{cm}}

\newcommand{\kpc}{\mathrm{kpc}}

\begin{document}

\begin{titlepage}

\begin{center}
\vspace*{-1cm}

\hfill RU-NHETC-09-15 \\
\vskip 1.0in
{\LARGE \bf Dark Matter Through the Neutrino Portal} \\
\vskip .1in

\vskip 0.5in
{\large Adam Falkowski},
{\large Jos\'e Juknevich},  {\rm and}
{\large Jessie Shelton}

\vskip 0.25in

{\em 
Department of Physics \\
Rutgers University \\
Piscataway, NJ 08854}

\vskip 0.75in

\end{center}

\baselineskip=16pt

\begin{abstract}

\noindent

We consider a model of dark matter whose most prominent signature is a
monochromatic flux of TeV neutrinos from the galactic center. As an
example of a general scenario, we consider a specific model where the dark matter is
a fermion in the adjoint representation of a hidden
$SU(N)$ gauge group that confines at GeV energies.  The absence
of light fermionic states in the dark sector ensures stability of dark
matter on cosmological time scales.  Dark matter couples to the
standard model via the neutrino portal, that is, the singlet operator
$H L$ constructed from the Higgs and lepton doublets, which is the
lowest dimensional fermionic singlet operator in the standard model.
This coupling prompts dark matter decay where the dominant decay
channel has one neutrino (and at least one dark glueball) in the final
state.  Other decay channels with charged standard model particles
involve more particles in the final state and are therefore 
suppressed by phase space.  In consequence, the standard indirect
detection signals like gamma-ray photons, antiprotons and positrons
are suppressed with respect to the neutrino signal. This coupling via 
the neutrino portal is most robustly constrained by Super-Kamiokande,
which restricts the dark matter lifetime
to be larger than $10^{25}$ seconds.  In the near future, the scenario
will be probed by the new generation of neutrino telescopes.  ANTARES
will be sensitive to a dark matter lifetime of order $10^{26}$
seconds, while IceCube/DeepCore can probe a lifetime as large as
$10^{27}$ seconds.

\end{abstract}

\end{titlepage}

\baselineskip=17pt

\newpage


\section{Introduction}

There is widespread hope that the next few years will see the
detection of a dark matter particle.  Signals of dark matter are
searched for obliquely in colliders, directly in underground detectors, and indirectly in
cosmic rays.  Of these three methods, the last one is {\it a priori}
the most challenging because backgrounds from astrophysical processes
are difficult to estimate.  Nevertheless, indirect detection is a
topic of great current interest due to rapid instrumental progress and
a constant flow of new and exciting data.  During the past year,
interest in indirect detection has been further amplified by
tantalizing signals from the PAMELA \cite{Adriani:2008zr}, ATIC
\cite{:2008zzr} and FERMI \cite{Abdo:2009zk} experiments, which may or
may not be hints of dark matter.

The main search channels for indirect dark matter detection are cosmic
ray photons, antiprotons, positrons, and neutrinos.  Certain features
of neutrinos make them an especially clean detection channel.
Neutrinos, like photons, have no electric charge and traverse our
galaxy along straight paths.  Therefore the flux of neutrinos observed
at the Earth is determined in a straightforward way by the production
mechanism and is free of the astrophysical uncertainties which plague
the propagation of charged particles.  Neutrinos have a further
advantage over photons in that astrophysical neutrino backgrounds are
far less severe. Detection of a large diffuse flux of astrophysical
neutrinos would thus constitute a smoking gun for dark matter.
 
In spite of these attractive features, neutrinos are seldom singled
out as a primary dark matter discovery channel.  Neutrinos are weakly
interacting and require large, dedicated experiments to detect.  It is
much easier to detect charged particles and photons.  Typical models
of dark matter encountered in the literature predict a cosmic neutrino
flux that is smaller than or comparable to the fluxes in other
observable channels.  Most often, dark matter dominantly annihilates
or decays into other standard model particles---$W$ bosons, for
instance, or tau leptons---which produce neutrinos when they
subsequently decay.  But then one would expect to first detect that
dark matter signal by observing antiprotons from $W$ decays or gamma
ray photons from $\pi^0$s in the tau lepton decay chain.  Annihilation
or decay to primary, hard neutrinos typically appears as only one of
several possible final states.  The $SU(2)$ gauge symmetry of the
standard model generically implies that charged particles are produced
at comparable rates, and those lead to more accessible signals. 
For these reasons measurement of the cosmic neutrino flux sometimes
provides useful constraints on dark matter models \cite{Tomer,buckley}, or may serve to
break a degeneracy among different models once a signal is detected in
another channel, but is not considered the main discovery channel.

In this paper we argue that one can construct dark matter models which 
predict a cosmic neutrino flux that is more prominent than
the antiproton, positron and gamma ray fluxes.  If such a model is
realized in nature, the indirect detection signal could first show up
in neutrino telescopes.  The neutrino flux from the galactic
center can well be larger than the atmospheric neutrino background,
leading to a sharp smoking-gun signature.  The current generation of
neutrino telescopes (IceCube and ANTARES) or the future one (KM3NET) may become the discovery machines.

What kind of set-up is required for dark matter to produce a large
neutrino flux?  It is plausible that the dark matter particle is a
perfect singlet under the standard model gauge interactions.  If such
is the case it can couple to the visible sector via gauge singlet
operators which can be constructed out of standard model fields.  The
lowest dimensional singlet operators are the Higgs mass $|H|^2$ and
the hypercharge field strength $B_{\mu\nu}$, both of mass dimension 2.
Dark matter coupling through the Higgs portal or via kinetic mixing
with hypercharge has been explored in numerous models.  However, if
the dark matter particle is a {\em fermion}, then the lowest dimensional
standard model operator it can couple to is the dimension 5/2 operator
\beq 
H L\equiv \epsilon_{ab} H ^{a} L ^ b = H ^ + e^-_L-H ^{0*}\nu_L 
        =G^+ e^-_L- \frac{1}{\sqrt 2}(v+h+iG ^ 0)\nu_L 
\eeq 
where $v = 246$ GeV
is the standard model Higgs vev.  We refer to this operator as {\em
the neutrino portal}.\footnote{The coupling through the neutrino
portal has been previously utilized in models of dark matter in a
somewhat different context \cite{Kaplan:2009ag}.}

Now, suppose that the fermionic dark matter particle $\lambda$ is a
part of a larger dark sector that is neutral under the standard model
gauge group.  Assume that the dark sector couples to the standard
model via the neutrino portal as
\beq 
\label{e.npc} 
\cl_{int} = O_{dark}(\lambda) (L H)
\eeq 
where $O_{dark}(\lambda)$ is a fermionic gauge singlet operator
constructed from the dark matter particle $\lambda$ and other fields
in the dark sector.  Suppose in addition that the dark sector contains
{\em bosonic} states that are lighter than $\lambda$.  If $\lambda$ is
the lightest {\em fermionic} state in the dark sector, it is stable
within the dark sector.  But in the presence of the coupling
\eref{npc} the dark matter particle is allowed to decay into fermionic
states in the standard model.  The leading dark matter decay channel
is then
\beq
\lambda \to {\rm (Dark)} + \nu 
\eeq 
which leads to a neutrino flux as the dominant signal from dark matter
decay!  Moreover, if there is only one dark sector particle in the
final state, the neutrino flux is monochromatic.  Of course, the
coupling in \eref{npc} also predicts other, subdominant decay
channels: $\lambda \to {\rm (Dark)} + (h\nu, W^\pm e^\mp, Z \nu)$,
which imply charged particle fluxes.  But these other decays have one
more particle in the final state, and are therefore suppressed by
phase space. The branching ratio for these states with on-shell bosons
depends on the multiplicity of dark particles in the final state and
the mass of the dark matter particle, and can be made as small as
$10^{-2}$.  Cosmic neutrinos then become the most prominent signal of
dark matter decay.

We present here a simple model with a strongly coupled dark sector
which explicitly realizes this idea.  The model contains an $SU(N)$
dark gauge group plus a massive Majorana fermion in the adjoint
representation.  With this low energy spectrum the dark gauge group is
asymptotically free for any $N$.  Therefore the dark color is confined
at low energies and the physical states are dark color singlets:
glueballs and glueballinos (bound states of the adjoint fermion and
gluons).  The lightest, ground state glueballino (in the following
simply called the glueballino) is our dark matter candidate.  Note
that the glueballino, being fermionic, cannot decay solely into
glueballs, even if it is heavier.  We assume that the dark confinement
scale is in the (sub)-GeV ballpark which sets the mass scale for the
glueballs, while the glueballino mass is set by the Majorana mass of
the adjoint fermion of order 1-5 TeV.  In the
absence of any coupling between the dark sector and the standard model
the glueballino would be perfectly stable.  Decay channels open up
once the adjoint fermion is coupled to the standard model via the
neutrino portal, that is, via the dimension six coupling ${c \over
\Lambda_N^2} {\rm Tr}[G_{\mu\nu }\ov \lambda] \sigma_{\mu\nu} (L
H)$.  For $\Lambda_N \sim 10^{14}-10^{15}$ GeV  and $c \sim 1/16 \pi^2$ (a
one-loop factor) the dark matter lifetime is much longer than the age of
the universe, while at the same time the decay rate is large enough to
yield observable signals.  The leading decay of dark matter is
the two-body decay into a dark gluon and a neutrino.  
This results in a flux of monochromatic neutrinos from the galactic 
center which can be  detected in the current generation of neutrino 
telescopes, ANTARES and IceCube.  Other indirect detection signals are 
suppressed, so that neutrino telescopes are singled out as the most 
sensitive probes of our scenario.  Moreover, the cross-section of our 
dark matter particle on nuclei is small, giving null predictions for 
direct detection experiments.

The paper is organized as follows.  In section 2 we introduce our
example model of the dark sector and discuss the spectrum after the dark gauge
group confines.  In section 3 we study the thermal relic abundance,
including the Sommerfeld effect and late annihilation of glueballinos.
Depending on assumptions about modeling nonperturbative contributions 
to the annihilation cross-section, the correct relic abundance may 
be obtained for dark matter masses in the range 1-10 TeV provided the
strong coupling scale is sufficiently small.
Besides the sizable theoretical uncertainties due to
modeling the strongly coupled effects at energies below the
confinement scale, the precise value of the mass is also sensitive to the number of degrees 
of freedom in the hidden sector -- the larger the number of colors, 
the smaller the required dark matter mass.   
In section 4 we compute the decay width and
the branching ratio for dark matter decays due to the coupling via the
neutrino portal.  We demonstrate that for the dark matter mass close
to a TeV the neutrino decay channel is by far dominant, while other
decay channels have branching ratios at a few percent level.  For
heavier dark matter, close to 5 TeV, the decay rate into charged
standard model particles becomes comparable to the decay rate into
neutrinos.  In section 5 we make a slight detour to explain why the
glueballs in the dark sector cannot be stable.  We enumerate various
options for their decay into the standard model final states and
indicate options for model building. 
Glueball decays may do lead to additional indirect detection signals which however are highly model dependent. 
For this reason we choose to focus on the more robust signals from the glueballino decays via the neutrino portal. 

In the later sections of the paper we turn to discussing observational
consequences of the general scenario where structure in the dark sector 
enables dark matter decay to yield a monochromatic neutrino flux as its 
primary signal. In section 6 we determine the
parameter space consistent with indirect detection
experiments.  The Super-Kamiokande bounds on the neutrino flux from
the galactic center set a lower bound on the dark matter lifetime, or
equivalently on the scale $\Lambda_N$ which suppresses the coupling
through the neutrino portal.  That scale turns out to lie in the
$10^{15}$ GeV ballpark---the typical scale for lepton violating
interactions.  We also study the current bounds from the antiproton,
positron and gamma ray fluxes. We conclude that even for the maximum
lifetime allowed by Super-Kamiokande the indirect bounds from other
detection channels are less constraining in a large portion of 
parameter space.  One exception are the bounds from antiprotons,
which are produced in subdominant dark matter decay channels with
standard model gauge and Higgs boson in the intermediate state.  These
bounds are comparable to those from Super-Kamiokande for dark matter
masses around 1 TeV, and when the dark matter mass is close to 5 TeV
(and thus the branching ratio for decays into charged particles
becomes large) the antiproton flux measured by the PAMELA experiment
implies a more severe constraint on the dark matter decay rate than
the Super-Kamiokande bounds.  However, one should note that
uncertainties involved in the propagation of antiprotons in our
galaxy are very large, and by changing the parameters of the
propagation model one can reduce the antiproton flux by a factor of
five, thus obliterating the antiproton bounds.  We therefore continue
to refer to the Super-Kamiokande bounds, which are comparatively free
of astrophysical uncertainties.  In section 7 we study the discovery
prospects in ANTARES and IceCube.  We compute the rate of signal
events in both detectors induced by the predicted neutrino flux from
the galactic center, and compare it to the background rate from
atmospheric neutrinos.  We conclude that neutrino telescopes have a
great potential to probe this scenario, and their sensitivity will
exceed that of Super-Kamiokande by 1-2 orders of magnitude.

\section{Model of Dark Sector}
\label{s.m}

In this section we construct an example of a dark matter model which
leads to an enhanced neutrino signal.  The model falls into the {\em
hidden valley} class \cite{Strassler:2006im}, where the hidden sector
contains fairly light states (here GeV scale) interacting very weakly
with the standard model through higher dimensional operators.  The
hidden sector here consists of a $SU(N)$ gauge group together with a
fermion in the adjoint representation, $\lambda^a$.  Incidentally,
this is also the field content of pure $\mc{N}= $1 super-Yang-Mills
theory, but we do not consider supersymmetry here: the fermion has a
TeV scale mass $m_\lambda$ and the interactions of the hidden sector
with the standard model are not supersymmetric.

\begin{figure}
\begin{center}
\includegraphics[width=10.0cm]{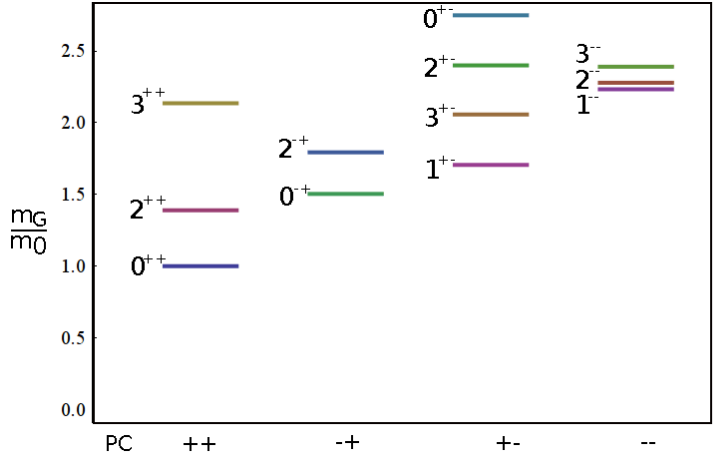}
\end{center}
\vspace{-0.75cm}
\begin{small}
\caption{Glueball spectrum for pure $SU(3)$ gauge theory, after \cite{Morningstar:1999rf}.  
Masses are shown in units of the lightest glueball mass $m_0$.
\label{fig:glue}}
\end{small}
\end{figure}

The dark gauge group is asymptotically free for any $N$ so that the
theory confines at low energies.  We assume that the confinement scale
is in the GeV ballpark, well below the adjoint fermion mass.  The
low-energy theory below a TeV contains only the dark gluons, and after
confinement the low-energy degrees of freedom are stable glueballs.
The spectrum of a pure Yang-Mills theory has been studied on the
lattice for the case $N = 3$ \cite{Morningstar:1999rf}.  These
studies indicate there are 12 stable glueballs in a pure $SU(3)$
theory, labeled by their $J ^{PC}$ quantum numbers.  The masses of the
glueballs are set by the strong coupling scale $\Lambda $.  The
lightest glueball is the $0 ^{++} $, with $m_{0++} = 4.2 r_0 ^{-1}$
where $r_0^{-1}$ is the ``force radius'' related to the confinement
scale $\Lambda$ (defined by the running coupling in the MS-bar scheme)
by  $\Lambda = 0.62 r_0 ^{-1}$ \cite{Gockeler:2005rv}.  Thus, the
lightest glueball mass
\beq 
m_{0++} \approx 7 \Lambda 
\eeq 
is appreciably larger than the confinement scale.  The spectrum of
glueball states for $N = 3$ is shown in figure~\ref{fig:glue}.  While
lattice results for the glueball spectrum are only available for
$N=3$, the masses of glueballs
do not vary appreciably with $N$ in the large-$N$ limit.  The spectrum
of the $N=3$ theory is thus a useful benchmark for more general cases.

In addition to glueballs, the spectrum of the dark sector contains
color singlet bound states of the adjoint fermion and gluons.
Borrowing from supersymmetric jargon we refer to the lightest of such
states as {\em the glueballino}.  The glueballino is stable under
decays within the dark sector, and is our dark matter candidate. 

\section{Relic Abundance of Dark Matter}
\label{s.th}

In the early universe at $T > 1$ TeV the adjoint fermions and the dark
gluons were in thermal equilibrium.  We assume that at even larger
temperatures the dark sector was in thermal equilibrium with the
standard model, but near a TeV the two sectors are already decoupled.
Below a TeV the adjoint fermions become non-relativistic and drop out
of thermal equilibrium, leaving a relic abundance that is in principle
determined by the fermion mass, the dark gauge coupling and the number
of dark colors.  The freezeout calculation of $\lambda$ abundance is
closely analogous to the well-studied problem of gluino dark matter
\cite{Baer:1998pg,Arvanitaki:2005fa}.  In the following we estimate
the range of fermion masses that are compatible with the observed dark
matter abundance.

\begin{figure}
\begin{center}
\psfrag{m}{$m_\lambda \:(\mathrm{TeV})$}
\psfrag{s}{$\Lambda \: (\mathrm{GeV})$}
\includegraphics[width=4in]{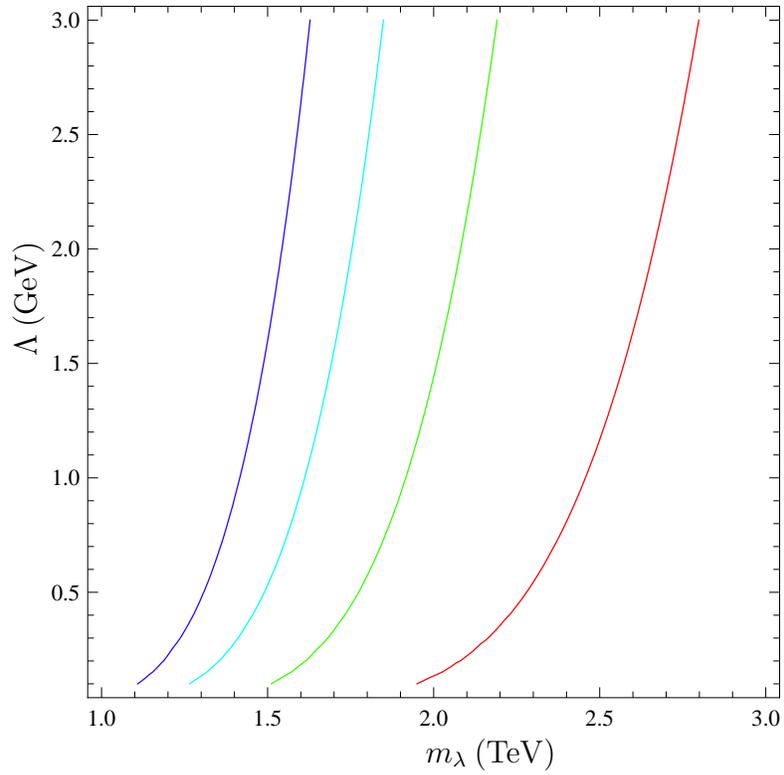}
\end{center}
\begin{small}
\caption{Contours of relic density $\Omega_\lambda = .23$ in the $m_\lambda $--$\Lambda/m_\lambda$ plane,
for $N = 3$ (red), $N = 4$ (green), $N = 5$ (cyan), $N = 6$ (blue).  Using the perturbative annihilation cross-section.
\label{fig:contour}}
\end{small}
\end{figure}

We first compute the relic abundance ignoring non-perturbative
effects, in which case the standard textbook procedure applies.  The
perturbative matrix element for the annihilation $\lambda\lambda\to
gg$ can be found in \cite{dhm}.
At low energies the thermally averaged annihilation cross-section is
given by
\beq
\la \sigma_{ann} v \ra  \approx \frac{3 N^2}{4 (N^2-1)} \frac{\pi\alpha_d^2}{m_\lambda^2}
\eeq 
where $\alpha_d$ is the dark coupling at the scale $m_\lambda$, given
in terms of the confinement scale and the number of dark colors by $\alpha_d = 6 \pi/11 N \log(m_\lambda/\Lambda)$.
Integrating the Boltzmann equation relates the annihilation cross section to  the relic abundance $Y_\infty$. 
The latter  can be translated into the dark matter density fraction today by $\Omega_\lambda \approx Y_\infty m_\lambda s_0/\rho_{crit}$
where $s_0 = 2889.2/\cm^3$, $\rho_{crit} = 3.95\times 10^{-47} \gev^4$; WMAP fixes $\Omega_\lambda$ to be $0.23$ \cite{wmap5}.
The final result strongly depends on the number of dark colors (the number of degrees of freedom in the hidden sector). 
Our results are plotted in Fig. \ref{fig:contour}. 


Nonperturbative contributions to the annihilation cross-section are
significant and introduce large uncertainties
\cite{Baer:1998pg,Arvanitaki:2005fa,Kang:2006yd}.  Below we estimate
the effects of Sommerfeld enhancement and of enhanced annihilation
after confinement.

The exchange of dark gluons between the nonrelativistic fermions leads
to an enhanced annihilation cross section at low velocity
\cite{Hisano:2003ec}.  The Coulomb interaction of two massive
particles takes place on a time scale set by the energy of the
scattering particles, $\tau_{\lambda\lambda} \sim ( m_\lambda v ^ 2)
^{-1} \sim 1/T$.  Meanwhile, the time scale for a massive particle to
interact with a gluon in the thermal bath, and therefore to randomize
its color, is set by its mean free path in the thermal bath.  This can
be estimated by $\tau_{\lambda g}=\ell = (n_g \langle\sigma_{g\lambda}
v\rangle)^{-1} \sim m_\lambda^2/T ^ 3$.  Since $\tau_{\lambda\lambda}
\ll \tau_{\lambda g}$, the two gluinos will remain in a state of
definite total color throughout the interaction, and the sign and
strength of their interaction will depend on that particular color
state.  We can solve the Schr\"odinger equation in the basis of
definite total color, and derive the Sommerfeld enhancement in each
color representation subspace $i$.  Two particles of equal mass moving
in a non-Abelian Coulomb potential have a scattering amplitude
enhanced from the plane wave result by the factor 
\beq
\label{eq:somm}
E(v_r) = \frac{ \frac{\pi \alpha_d \xi}{v_r}}{1-e ^{-\pi \alpha_d \xi/v_r}} ,
\eeq
where $v_r$ is the {\em relative} velocity of the scattering
particles, and $\xi$ depends on the quadratic Casimir operators of the
single-particle representations $r_1$ and $r_2$ and the total
representation $r$,
\beq
\xi =-\frac{1}{4} (C_2 (r) -C_2 (r_1)-C_2 (r_2)).
\eeq
The annihilation cross-section during freeze-out is then multiplied 
by the factor
\beq
\label{eq:prescription}
E_{ann}(v_r; \{ r \}) = \frac{1}{\mathrm{dim} (r_1)\mathrm{dim} (r_2)}\sum_{i} \mathrm{dim} (r_i) E_i (v_r),
\eeq
which enhances annihilation in attractive channels and exponentially
suppresses annihilation in repulsive channels.

\begin{figure}
\begin{center}
\psfrag{m}{$m_\lambda \:(\mathrm{TeV})$}
\psfrag{Omega}{$\Omega_\lambda $}
\includegraphics[width=4in]{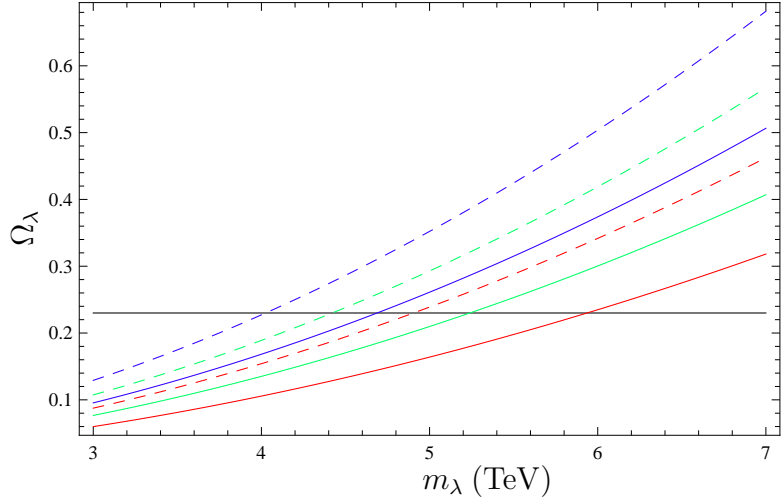}
\end{center}
\vspace{-0.75cm}
\begin{small}
\caption{Relic density $\Omega_\lambda$ as a function of $m_\lambda $, with (dashed)
and without (solid) Sommerfeld enhancement, for $N = 2$. The solid line indicates $\Omega_\lambda= .23$. 
Relic densities are shown for $m_\lambda/\Lambda = 0.5 \times 10 ^ 3$ (red), $m_\lambda/\Lambda = 1 \times 10 ^ 3$ (green), 
and $m_\lambda/\Lambda = 2 \times 10 ^ 3$ (blue).
\label{fig:somm}}
\end{small}
\end{figure}

As the gluinos are identical particles, we need to ensure that the
total wave function is antisymmetric.  Annihilation proceeds
dominantly in the $s$-wave, where the wave function is spatially
symmetric, and therefore the net color and spin wave function must be
antisymmetric.  Thus for $N = 2$, the total enhancement factor $E (v)
$ is explicitly
\beq
E (v_r) = \frac{1}{15}\left (  \frac{ \frac{\pi \alpha_d}{v_r}}{1-e ^{-\pi\alpha_d/v_r}}
        + 9\, \frac{ \frac{\pi \alpha_d}{2v_r}}{1-e ^{-\pi\alpha_d/2v_r}}
       - 5\, \frac{ \frac{\pi \alpha_d}{2v_r}}{1-e ^{\pi\alpha_d/2v_r}} \right)
\eeq
Incorporating the Sommerfeld enhancement has an order $30\%$ effect on
the mass of the dark matter particle relative to perturbative
freezeout, as demonstrated in figure~\ref{fig:somm}.

More difficult to estimate are the effects of the increased annihilation 
cross-sections after confinement, when the $\lambda ^ a $ have hadronized
\cite{Baer:1998pg,Arvanitaki:2005fa,Kang:2006yd}.  The heavy fermion
$\lambda$ is localized within a region of order its de Broglie
wavelength, $1/(m_\lambda v) $, while the total radius of the
glueballino is of order $1/\Lambda $. While the {\em total}
interaction cross-section between hadronized adjoints should be
geometric, $\sigma_{tot} \sim \pi/ \Lambda ^ 2$, the annihilation
cross-section should be significantly smaller.  Estimating the
annihilation cross-section is a subtle problem.  The core of the
question is whether or not two hadronized $\lambda$'s in a state
with large angular momentum are kinematically able to radiate away
that angular momentum and annihilate.  Our hidden sector model differs
from QCD in that there are no light fundamental quarks, and hence the
mass gap in our hidden sector is larger compared to the strong
coupling scale than the mass gap in QCD.  We expect that this
suppresses the annihilation cross-section in this model, relative to
(e.g.) gluino freezeout, as the larger mass gap makes it more
difficult to radiate away angular momentum.

We therefore model the late-time annihilation cross-section by taking
it to saturate the $s$-wave unitarity bound,
\beq
\label{eq:np}
\sigma_{NP} \approx \frac{16\pi}{m_{\lambda}^2 v_r^ 2} .
\eeq
We expect that this estimate gives an upper bound on the cross-section, 
and therefore a lower bound on the relic abundance. 
This cross-section yields efficient annihilation of hadronized adjoints, 
lowering the relic density by roughly an order of magnitude relative to 
the perturbative result.  It is therefore difficult to obtain
the relic abundance $\Omega_\lambda = 0.23$ without going to 
corners of parameter space, either large $N$ or $m$, or small $\Lambda$.

As we will explain in the following section, the most interesting choice of dark matter masses from the point of view of neutrino signals is the range $1$--$5 \tev$. This range is consistent with the perturbative computation of the relic abundance. However, if the non-perturbative effects enhancing the annihilation cross-section are close to our estimate \er{np}, this mass range requires either sub-GeV $\Lambda$  or a very large ($N>20$) number of dark colors. 
Building a complete model of the dark sector with $\Lambda$  significantly less than a GeV requires some additional structure in the dark sector to be consistent with all cosmological constraints, as we discuss in section~\ref{sec:poof}.

\section{Dark Matter Decay}
\label{s.dmd}

The low-energy lagrangian that governs the hidden sector has an
accidental $\Z_2$ symmetry acting as $\lambda \to -\lambda $ which
ensures stability of dark matter.  However, general interaction terms
do not have to respect this $\Z_2$ and may lead to dark matter decay.
In this section we consider the coupling of dark matter to the
standard model via the neutrino portal and study possible decay
channels.

The operator $\lambda^a \sigma^{\mu \nu} G_{\mu\nu}^a$ transforms as a
fermion under the Lorentz group and as a singlet under both the dark
and the standard model gauge groups.  Therefore it has the right
quantum numbers to couple to the standard model through the neutrino
portal.  The interaction takes the form
\beq
\label{eq:portal}
\mcL_{int} = \frac{c}{\Lambda_N^2} \lambda^a\sigma^{\mu\nu} G^a_{\mu\nu} H L +\hc.
\eeq
Here $L$ is a left-handed weak $SU(2)$ lepton doublet; for
definiteness let as assume that it is the first generation doublet,
$L = (\nu_L,e_L)$, while the couplings to the second and third
generations are suppressed by a larger scale.  From the point of view
of neutrino detection it does not matter which generation has the
dominant dominant coupling to dark matter: the neutrino oscillations
will make sure that all 3 neutrino flavors are equally populated in
the neutrino flux from dark matter decay.

The interaction \er{portal} breaks the accidental $\Z_2$
$\lambda$-parity and allows the dark matter to decay.  It is a
dimension six non-renormalizable operator, but it can be easily
obtained from a renormalizable UV theory as follows.  Let us introduce
right-handed neutrinos $N_i$ which have the usual Standard Model
couplings.  In addition, introduce a pair of fundamentals in the
hidden sector: a scalar $Q$ and a (Dirac) fermion $\chi$, which have
masses equal to $\Lambda_N$.  The lagrangian
\beq
\mcL = \mcL_{kin} - \lambda ^ a (  Q^\dag t ^ a \chi) -c_i  Q^\dag \chi N_i -  Q^\dag Q |H|^2 -Y_{ij}^N N_i H L_j +\hc 
\eeq
contains the most general gauge-invariant renormalizable operators
which can be built from the fields in the model.  Integrating out the
heavy $Q$, $\chi$, and $N_i$ at one loop gives the effective
interactions
\beq
\mcL_{int} = \frac{c}{\Lambda_N^2}\lambda ^ a\sigma ^{\mu\nu} G ^ a_{\mu\nu} H L_L + \frac{c_2 m_\lambda}{\Lambda_N^2} \lambda^a \lambda^a | H| ^ 2+ 
              \frac{c_3}{\Lambda_N^2} G ^ a_{\mu\nu} G ^ {a \mu\nu} | H | ^ 2 +\hc. 
\eeq
The coefficients $c$, $c_2$, and $c_3$ are of order $1/16\pi^2$.  The
latter two interactions have negligible effect on phenomenology, but
the first is the desired coupling of dark matter to the standard model
via the neutrino portal.  If the scale $\Lambda_N$ is of order
$10^{14}-10^{15}$ GeV---the typical scale for lepton number violating
interactions---then the dark matter lifetime is of order $10^{25}$
seconds.  Such a dark matter particle is stable on cosmological time
scales and, at the same time, the decay rate may be large enough to
allow for detection of dark matter decay products.



The perturbative vertex given by \er{portal} yields two distinct
partonic contributions to dark matter decay: first, the $1\to 3$
process $\lambda\to g \, H\, L$ with a (hidden) gluon in the final
state, and second, the $2\to 2$ process $\lambda\, g\to H\,L$, with a
(hidden) gluon in the initial state.  Both partonic processes are part
of the full gauge invariant amplitude for the decay of the dark matter
particle, but, as we will see, the first process is the dominant
contribution.  The main contribution to dark matter decay is then the
decay of the heavy adjoint $\lambda^a$ via \er{portal} to two-body and
three-body final states with a hidden gluon plus standard model
neutrinos, gauge and Higgs bosons.  The hard gluon $g$ from the
perturbative decay and the remnants of the brown muck that made up the
glueballino go on to hadronize into some number of hidden glueballs.

We first consider the two-body decay with one neutrino and one dark
gluon in the final state.  The interaction vertex is obtained from
\er{portal} by replacing the Higgs field by its vacuum expectation
value.
The partial decay width is given by 
\beq
\label{eq:2bodyG}
\Gamma (\lambda\to g\nu) \equiv \Gamma_{2b} = \frac{c^2}{32 \pi}\, \frac{m_\lambda^3 v^2}{\Lambda_N^4}.
\eeq
The CP conjugate decay to anti-neutrinos has the same decay width. 

Next, consider the three-body decay to gluon, neutrino, and on-shell
Higgs boson.  The partial width for this process is
\beq
\Gamma (\lambda\to g h \nu) = \frac{c^2}{(16\pi)^3}\frac{m_\lambda^5}{\Lambda_N^4} 
   \frac{1}{3}\left( 3 + 44r - 36 r^ 2 - 12 r^3 + r^4 + 12 r (2+3 r)\ln r \right),
\eeq
where  $r = m_h^2/m_\lambda^2 \ll 1$. Again, the CP conjugate decay to anti-neutrinos occurs with the same rate.
In the limit $r \to 0$ 
\beq
\Gamma (\lambda\to g h \nu) =  \Gamma (\lambda\to g h \bar \nu)  \approx    \frac{c^2}{(16\pi)^3} \frac{m_\lambda^5}{\Lambda_N^4}  \equiv \Gamma_{3b}. 
\eeq 
In addition, there are three-body decay modes to (i) gluon, electron,
and $W$ boson; (ii) gluon, neutrino and $Z$ boson.
In the limit $m_{W,Z}/m_\lambda \to 0$ these partial widths satisfy the relation  
\beq 
\Gamma (\lambda\to g Z \nu) = \Gamma (\lambda\to g Z \bar \nu) = {1 \over 2}\Gamma (\lambda\to g W^+ e^-) = {1 \over 2}\Gamma (\lambda\to g W^- e^+)  = \Gamma_{3b} 
\eeq 
Ignoring gauge and Higgs masses, the dark matter lifetime is approximately  
\beq
\tau_\lambda \approx {1 \over 2(\Gamma_{2b} + 4 \Gamma_{3b})},  
\eeq  
while the ratio of two- and three-body partial widths is 
\beq
\label{eq:2to3}
{\Gamma_{3b} \over \Gamma_{2b}} \approx   {1 \over 128 \pi^2 } {m_\lambda^2 \over v^2} \approx 0.01 \left (m_\lambda \over \tev \right )^2
\eeq  
Varying $m_{\lambda}$ in the range 1-5 TeV, the three-to-two-body ratio varies from 1 percent to 30
percent.  Therefore the neutrino flux from dark matter decay dominates
over charged particle fluxes, especially when the glueballino is not
much heavier than 1 TeV.

Apart from the process where the adjoint fermion decays into a dark gluon and standard model fields one can also  consider the annihillation 
of the fermion with a valence gluon. 
We argue however that this gives a negligible contribution to glueballino decay.  
First, it is readily apparent that the rate for the gluino to
completely annihilate with its cloud of gluons is vanishingly small.
In order for the glueballino to annihilate into the vacuum, all of the
momentum distributed among the brown muck must be concentrated in a
single partonic gluon which then annihilates with the heavy fermion
$\lambda$, or in other words, all of the momentum carried by the
gluons must be entirely localized within the Compton wavelength of the
heavy fermion $\lambda$.  The probability for this to occur is
negligibly small, of order $(\Lambda/m_\lambda) ^ 3$ at most.
Moreover, the probability for the operator $\lambda ^ a \sigma
^{\mu\nu} G ^ a_{\mu\nu}$ to connect the initial state glueballino to
a single glueball at rest is suppressed.  This process can be thought
of as an annihilation of the partonic $\lambda^a$ with a coherent sum
of soft gluons inside the hadron.  The residual soft gluons after
annihilation must carry enough energy to reform into a stable
glueball.  As the glueballs have masses of a factor of a several
larger than the strong coupling scale, an annihilation of this form is
kinematically forbidden unless the $\lambda^ a $ has a virtuality
likewise several times larger than the strong coupling scale.
Computing the cross-section for $\lambda ^ a+g\to h \nu$ and taking
$p_g \sim \Lambda $, we can estimate
\beq
\sigma \sim \frac{c ^ 2 }{ \Lambda_N^ 4} \frac{ m_\lambda \Lambda (1-r) ^ 2}{24\pi}.
\eeq
Strictly, as the gluon participating in the interaction is soft we
need to integrate this expression over the field configuration of the
glueballino.  To obtain an estimate for the annihilation rate we
simply estimate the result of this integral as $\Lambda$.  This yields
an estimate for the rate
\beq
\Gamma \sim \frac{c ^ 2 }{ \Lambda_N^ 4} \frac{ m_\lambda \Lambda^ 2 (1-r) ^ 2}{24\pi}:
\eeq
this is suppressed relative to the two-body decay rate \er{2bodyG} by
a factor of $(\Lambda/v) ^ 2$.  However the rate for the full process
should be suppressed even further, as the gluino must have emitted a
hard gluon before participating in the annihilation in order to be
sufficiently off-shell.  This emission suppresses the annihilation
process by an additional factor of $\alpha_d$. In the following, we will
neglect this glueballino decay mode in comparison with the two-and
three-body decay modes coming from the decay of the gluino alone.

\section{Interlude: Glueball Decays}
\label{sec:poof}

The hidden sector glueballs cannot be stable on cosmological time
scales as otherwise they would overclose the universe.  In this section we discuss the thermal relic
abundance of glueballs and present several scenarios for their decay.
We will show that there are multiple ways to eliminate the
overabundance of glueballs, and will indicate the observable signals
associated with each scenario.  At the end of the day it will be clear
that the signals and resulting constraints are very model-dependent,
in contrast to the sharp predictions from two- and three-body decays
via the neutrino portal discussed in \sref{dmd}.  
Therefore, when discussing signals and constraints in the following of this paper we will  focus on the more robust 
signals of the standard model particles coming directly from the neutrino  portal operator $\lambda^a G^a_{\mu\nu}\sigma ^{\mu\nu} H L $.
The goal of this section is to point out and characterize a range of options for
addressing the glueball problem, and demonstrate that solutions are
possible.  Readers less interested in model building and more
interested in model-independent astrophysical signatures of the
neutrino portal are encouraged to jump directly to the next section.

First, we demonstrate why the glueballs must decay.  Treating hidden
sector confinement as adiabatic, the energy and entropy in the thermal
gluon bath are wholly transferred to glueballs after confinement. If
the glueballs do not decay, the present-day energy density in
glueballs is then given in terms of the present-day photon temperature
$T_{0}$ by
\beq
\label{eq:gluefoo}
\rho_{glue} (t_0) \approx \frac{\pi ^ 2}{30} \; \Lambda \; T ^ 3_{0} \frac{g_{*0} \;g_{*D}}{g_{*SM}}\, ,
\eeq
where $g_{*,0}$ is the effective number of relativistic standard model 
degrees of freedom today, and $g_{*,D,SM}$ are the effective number of 
relativistic degrees of freedom in the dark sector and in the standard 
model when the two sectors were in thermal equilibrium. 
Avoiding overclosure then requires the strong coupling scale $\Lambda$ to be smaller than an eV,
\beq
(N ^ 2-1) \frac{\Lambda}{\ev} \lsim 1.
\eeq
However, the strong coupling scale $\Lambda$ is constrained by
astrophysical bounds on the dark matter self-interaction
cross-section.  Analyses of galactic cluster dynamics combine to yield
a bound on the total dark matter self-interaction cross-section
\cite{Dave:2000ar,Gnedin:2000ea,Markevitch:2001ri,Randall:2007ph},
\beq
\label{eq:xsec}
\frac{\sigma}{m} \lsim  0.3 - 1 \;\mathrm{cm} ^ 2 \mathrm{g} ^{-1} \approx 1400 - 4600 \gev^{-3} .
\eeq
The most stringent such astrophysical bound comes from observations of
the Bullet Cluster, 1E0657-56
\cite{Markevitch:2001ri,Randall:2007ph}.  In our model, the {\it
total} self-interaction cross-section for the glueballino is
geometric, $\sigma_{tot} \sim \pi/ \Lambda ^ 2$.  With $\Lambda \sim
\ev$, the bound \er{xsec} requires the mass of the glueballino to be
very large, $m_\lambda \gsim 10^ 9\gev$, which would be incompatible with thermal relic abundance. 

As the glueball masses are intrinsically tied to the confinement
scale, there is no way to escape this problem.  The glueballs must
therefore decay.  There are then three possibilities: the glueballs
can decay within their own sector; they can decay to the standard
model; or finally, they can decay to yet another sector.  Adding light
flavors to the hidden sector gauge theory destabilizes the glueballs
to hidden sector mesons and opens an interesting range of
possibilities for model building, which we discuss below.

Another option, which we will pursue here, is to add additional
degrees of freedom (``mediators") which couple the hidden and visible
sector, enabling glueballs to decay to standard model final states.
Dark matter decays then yield a model-dependent amount of visible
standard model particles from glueball decays in addition to the
primary gluino decay, which must be taken into account.  In order to 
avoid cosmological constraints, it is simplest to ask that the
lifetime of at least one species of glueball be less than one second,
so that the glueballs can safely decay without a trace before
nucleosynthesis.  For definiteness, we will consider two simple models 
where the mediators couple to the standard model Higgs boson, thereby 
enabling the $0^{++}$ glueball to decay rapidly through the dimension 
6 Higgs portal operator to pairs of standard model fermions.  In the 
first model, we introduce a standard model singlet scalar field $\phi$ 
transforming as a fundamental under the dark gauge group, with the 
interaction
\beq
\label{eq:scalar}
\mc{L}_{int} = y |\phi| ^ 2 |H|^2 .
\eeq
In the second model, we introduce a pair of Dirac fermions, $K_L$ and
$P_R$, which are fundamentals under the dark gauge group and have the
standard model quantum numbers of a left-handed lepton doublet and a
right-handed charged lepton, respectively.  The interactions of the
new fermions $K_L$ and $P_R$ we take to be $C$- and $P $-preserving,
for simplicity:
\beq
\label{eq:fermion}
{\cal L} =  {\cal L}_{\mathrm{kin}}+\left(y \bar P_R H K_L  +h.c.\right) .
\eeq
Integrating out the mediators induces the dimension six interaction
\beq
\label{eq:d6}
{\cal L}_{eff} =  \frac{ g^2 c }{4 \pi^2 M^2}  \, H^\dagger H\, G^a_{\mu\nu}G^{a\mu\nu}.
\eeq
Here $M $ is the mass scale of the mediators.  The coefficient $c$ is
$y ^ 2/12 $ for the scalar mediator model of equation~\er{scalar}, and
$2 y ^ 2/3$ for the fermion mediator model of equation~\er{fermion}.

The $0 ^{++}$ glueball decays through the effective interaction
\er{d6} with branching fractions dictated by the couplings of the
standard model Higgs.  For the parameter range we consider, the
$0^{++}$ will decay to pairs of standard model fermions through an
off-shell Higgs. The width for this decay is given by
\beq
\label{eq:0width}
\Gamma_{0^{++}\rightarrow f\bar f} = \left(\frac{c v F_{0^+}^S}{16 \pi^ 2 M^2 (m_h^2-m_0^2)} \right)^2 \Gamma^{SM}_{h \rightarrow f\bar f}(m_0^2)
\eeq 
Here $\Gamma^{SM}_{h \rightarrow f\bar f}(m_{0}^2)$ denotes the
partial width of $h \rightarrow f\bar f$ for a standard model Higgs
boson with mass $m_{0}$, and $F_{0^+}^S$ is the $0 ^{++}$ decay
constant.  We will use the lattice result for $F_{0^+}^S$
\cite{Chen:2005mg} ,
\beq
F_0^S \equiv g^2 \langle 0|G^a_{\mu\nu}G^{a\mu\nu}|0^{++}\rangle = 6.12 \, m_0^3 .
\eeq 

If the $0 ^{++} $ is heavy enough to decay to $b\bar b $ pairs, a lifetime less
than one second is readily accomplished with mediator mass scales in
the range of tens to hundreds of TeV.  Lighter glueballs, with masses
in the range $2 m_\tau < m_{0 ^{++}} <2 m_b $, are also viable if the
mediator mass scales are in the range of a few to tens of TeV.
Glueballs light enough to decay dominantly to muon pairs are interesting,
as this mass range appears best compatible with obtaining the correct relic
density for the heavy hadronized adjoint, as well as with possible
visible signals.  However, realizing a sufficiently short lifetime for the $0 ^{++}
$ in this case requires the mediators to be so light that they lie outside the
range of validity of our calculations.  Models with light standard
model-charged mediators are ruled out, but a light scalar
mixing with the standard model Higgs is entirely possible, and seems
like an interesting avenue to pursue.

In the case of a scalar mediator, the only interactions induced upon
integrating out the heavy mediators are of the form
\beq
\mc{L}_{eff} = G^a_{\mu\nu}G^{a\mu\nu}\; \sum_k \frac{c_k }{M ^{2+k}} (H^\dagger H)^ k .
\eeq
The Lorentz decomposition of the operator $G^a_{\mu\nu}G^{a\mu\nu}$
contains only the $0 ^{++} $.  Therefore only the $0 ^{++}$ can decay
directly via the Higgs portal to standard model particles.  Other
glueballs can decay through the $0 ^{++}$ channel, e.g., $2^{++}\to (0 ^{++}) ^* \, 0 ^{++} \to
f\bar f\, 0 ^{++}$.  These three-body decay rates are highly suppressed
relative to the $0 ^{++}$ decay rate due to the smaller available phase
space.  The only glueballs which cannot decay in this manner are the
lightest states in the $P$ and $C$-odd sectors, the $0^{-+}$ and
$1^{+-}$.  Without explicit $P$ or $C$ violation, no dimension 6
operator permits the $0^{-+}$ and $1^{+-}$ states to decay into
standard model fermions.

In the case of bifundamental mediators, loops of the heavy particles
also induce dimension 8 effective interactions coupling the hidden
gluons to standard model gauge bosons,
\beq
\label{eq:d8}
\mc{L}_{eff} = c_1 \frac{g_i^2 g_v^2}{M^4}\,\tr F_i^2 \,\tr \,G^2 + c_2  \frac{g_1 g_v^3}{M^4}\, F_1 \,\tr \, G^3 ,
\eq
considered in detail in~\cite{Juknevich:2009ji} in the context of
hidden valley signals at the LHC.  Here $F$ $(G)$ is the field
strength tensor for standard model (dark sector) gauge fields. The
field strength tensors are contracted according to different
irreducible representations of the Lorentz group which for the sake of
brevity are not shown here. These dimension 8 operators allow all
the hidden sector glueballs to decay, either directly to standard model gauge
boson pairs (the $0^{++}$, $2^{++}$, $0^{-+}$, $2^{-+}$) or
radiatively to another glueball  via the emission of a
photon (all others).

For a glueball which can decay either through
radiative photon emission or through an off-shell Higgs, the situation
is rather involved. The branching ratios of these two modes depend
critically on the mass scale $M$ of the mediators and the
nonperturbative transition matrix elements. 
The ratio of the partial widths can be written parametrically as
 \beq
 \frac{\Gamma_{A\to C \gamma}}{\Gamma_{A\to B ff}} \approx
 \left(\frac{m_H}{y M}\right)^4
                 \left(\frac{m_A}{m_f}\right)^2 \alpha \alpha_d \chi^2
 \frac{1}{I\left(\frac{m_B}{m_A},\frac{m_f}{m_A}\right)}
 \eeq
 where $\chi = M^4(1/m_K^4-1/m_P^4) $ and $\alpha$ is the fine structure
 constant.  The dimensionless phase-space  function $I\left(a,b\right)$
  is given by
 \beq
 I(a,b)= \int_{4b^2}^{(1-a)^2} dx\, x \, \left(g(a^2,x;
 1)\right)^{L+1/2} \left(g(b^2,b^2;x)\right)^{3/2}.
 \eeq
 where $g(x,y;z)= (1-x/z-y/z)^2-4x y/z^2$ and $L$ is the relative
orbital angular  momentum of $A$ and $B$.
 Given the masses in Fig. \ref{fig:glue}, $I \approx10^{-6}-10^{-5}$.  For $\chi \sim
0.1-0.5$, which is a conservative choice if the mass splittings between
the mediator fields are not especially large, the larger suppression
of dimension eight operators tends to suppress the radiative decays in
favor of the three-body decays.  However, for small enough $M$, the
radiative decays can become competitive. Indeed, a simple estimate
suggests that for $M\gsim 10\tev$, the dimension 6 operator dominates,
while for mediators within the reach of the LHC radiative and
three-body rates can be comparable.  If the off-shell Higgs needs to decay to
muon pairs, the small muon Yukawa coupling will additionally suppress the
three-body rates, making the radiative decays dominate.

To summarize, glueball decays contribute $b$ quarks or $\tau$ leptons
to the visible signals of dark matter decay, as well as possible
photons (both singly and in pairs) if the mediators are charged under
$SU(2)_L\times U(1)_Y$.  (Decays into $W$ and $Z$ bosons are
kinematically impossible.)  Colored mediators would also allow the
glueballs to decay into standard model gluon pairs, but in order to
minimize the antiprotons coming from glueball decays we have taken
the mediators to be color singlets.  Moreover, as most of the mass
splittings between glueballs are smaller than $m_0$, if the $0 ^ {++}
$ is near the threshold to decay to $b\bar b$ pairs, then kinematics
may prevent $b\bar b$ pairs from being produced in three body decays
of heavier glueballs.  A typical final visible state would then be of
the form $b\bar b c\bar c,$ $b\bar b \tau^ +\tau ^-$ and so on.  A
similar story holds for $0 ^ {++} $ decaying to tau pairs:
higher-mass glueballs could yield pion and muon pairs in addition to
the tau pairs coming from the decay of the terminal $0 ^ {++} $.
Glueballs which decay mainly into muon or pion pairs might
be obtained with additional model building in the Higgs sector.

While it is possible to work out glueball branching fractions 
for a given mediation model with the help of lattice data for $N=3$, 
it is considerably more difficult to arrive at the 
full spectrum of visible standard model particles produced in a single 
glueballino decay.  The injection spectrum $dN_{SM}/dE$ of visible 
standard model particles depends on the number and spectra of
glueballs initially produced in the glueballino decay.  In order to
make crisp predictions about indirect detection signals coming
from dark matter decay, we therefore need to understand the process of
fragmentation in pure Yang-Mills theory.  Unfortunately, this is a
situation where neither phenomenological examples from QCD nor data
from the lattice can be of help. In the absence of light fundamentals,
the color tube connecting hard partons cannot break. Glueball formation in
a pure Yang-Mills theory occurs heuristically through the crossing of
the tube onto itself, a process qualitatively different from
fragmentation in QCD.

One simple approach to estimating the relative abundances of different
glueball species is a thermal model.  In such a model, the ratio of
the multiplicity of glueballs is given by
\beq
\frac{N_J}{N_{0}} = (2J+1)\left(\frac{m_J}{m_0}\right)^{3/2} \exp^{-(m_J-m_0)/T}
\eeq
where $m_J$ $(N_J)$ is the mass (multiplicity) of a glueball with spin
$J$ .  The effective temperature $T$ can be taken as the center of mass
energy of the colored system immediately following the decay of the
heavy adjoint fermion, $T\sim \sqrt{\Lambda m_\lambda}$. With
$m_\lambda \gg \Lambda$, the thermal model predicts glueball democracy
in the final state: for $m_\lambda = 1\tev$ and $\Lambda = 1\gev $,
the heaviest of the stable glueballs, the $0^{+-} $, is produced
two-thirds as frequently as the $0^{++}$ .  Unfortunately, the
applicability of the thermal model to pure Yang-Mills theory is
unclear.  In particular, excitations of the flux tube (such as can
easily be produced from gluon showering in the final state) might
enhance the production of the higher-spin states above the thermal
estimate.  Moreover, the thermal model is not well-suited to address a key
question for indirect detection: what is the average {\em total}
number of glueballs per decay?

To arrive at an injection spectrum $dN_{SM}/dE$ which can be compared 
to indirect detection data, we need to specify not only a particular 
model for glueball decay but also a model for fragmentation in the hidden 
sector.  While it appears possible to build interesting models which 
have the potential to avoid overproduction of antiprotons, it is clear that
not only is there significant freedom in the  detailed choice of model,
but also that a reasonable model of pure-glue fragmentation is a necessary
ingredient in quantifying the predictions  of such models for indirect detection
experiments.

\subsection{Model Building with Light Hidden Flavor}

Adding light ($m_q \lsim\Lambda $) fundamental flavors to the hidden
sector destabilizes the glueballs to hidden mesons, without
destabilizing the adjoint fermion $\lambda$.  As $\lambda$ is
fermionic, it must decay to an odd number of hidden quarks and
anti-quarks.  But such a final state carries non-zero hidden-sector
baryon number.  The decay of $\lambda$ to light flavor therefore
violates baryon number, and is therefore forbidden in the absence of
hidden sector baryon number violation.

The meson masses depend on the input light quark masses as well as on
the strong coupling scale, and can be tuned independently of the other
parameters of the theory.  Avoiding overclosure requires that the
hidden mesons have sub-eV masses, giving rise to a large but
technically natural hierarchy, $m_q\ll\Lambda\ll m_\lambda $.
However, the existence of such extremely light mesons leads to
concerns that inelastic glueballino scattering $\Psi \Psi\to \Psi \Psi
+ n \pi_v$ may exceed observational bounds on dissipative dark matter
interactions. Adding such extremely light mesons will also
dramatically increase the late-time glueballino annihilation
cross-section, likely to a problematic extent.

More attractive model building possibilities are realized by keeping
$m_q\sim\Lambda$ and allowing the hidden sector mesons to decay to 
the standard model.  This can be realized in multiple ways.  One 
option is to introduce a $Z'$ coupling to both hidden flavor and 
visible standard model fermions \cite{Strassler:2006im}.  This 
allows at least one hidden sector meson to decay to a pair of 
standard model fermions, with a preference for heavier fermions 
due to helicity suppression.  The main challenge is then obtaining a 
short enough lifetime for the shortest-lived hidden sector meson while 
keeping $m_q\sim\Lambda\sim m_{\pi_v}$ at the GeV scale.  For models of the form 
considered in \cite{Strassler:2006im}, lifetimes of less than a second 
can be achieved for $2m_\tau< m_q\sim\Lambda $, with the mediator mass
in the range of a few to tens of TeV.  Decay to muon pairs can only be
realized for a light mediator and $m_{\pi_v} $ close to the $\tau$ threshold.

Models with light flavor have one especially nice feature: they 
allow for more concrete predictions for the visible signals from 
dark matter decay, as with the addition of light flavor it is 
possible to adapt models of fragmentation in QCD.  This then allows
definite predictions for the spectra and multiplicities of the hidden
sector hadrons produced in dark matter decay, which in turn allows
detailed comparison to astrophysical data.  While this is appealing,
it comes at a cost: the existence of light degrees of freedom in the 
hidden sector exacerbates the uncertainties in evaluating the 
annihilation cross-section for glueballinos, and hence in the acceptable 
mass range for the dark matter.  Nonetheless, there are interesting avenues 
for model building here.

\section{Constraints from indirect detection}

We turn to discussing the constraints on our neutrino portal scenario from various
indirect detection experiments.  The glueballino decaying via the
operator of \eref{npc} produces dominantly neutrinos, but also Higgs
particles, $W$ and $Z$ bosons, and electrons and positrons.  These
fluxes of standard model particles produced directly from the neutrino 
portal operator are relatively model-independent: the relative normalization of
the two- and three-body decay modes depends only on the kinematics of the
decay.  Bounds on these decay modes therefore constrain the general scenario
where structure in the dark sector enables dark matter decaying through 
the neutrino portal to yield a primary flux of monochromatic neutrinos.  We
compute the fluxes of neutrinos, antiprotons, positrons, and
gamma rays observed at the Earth for standard model particles produced 
in the decay of the heavy dark matter particle, and determine the allowed parameter
space of our model.

Throughout this section we assume the dark matter distribution in our
galaxy is described by the Einasto profile
\beq 
\label{e.ep} 
\rho_{DM}(r) = \rho_0 \exp \left [-2{(r/r_s)^\alpha - 1 \over \alpha} \right ]
\eeq 
where $r$ is the distance from the center of our galaxy, and we use
the values $\alpha =0.17$ and $r_s = 20$ kpc \cite{einasto}.  The
normalization constant is $\rho_0 = 0.06\gev/{\rm cm}^3$ which gives
the local dark matter energy density $\rho_{\sun} \equiv
\rho_{DM}(r_{\sun}) = 0.3 \gev/{\rm cm}^3$. Here $r_{\sun} = 8.5$ kpc
is the distance of the Solar System from the galactic center.  Because
for decaying dark matter the signal depends on $\rho_{DM}$ (rather
than $\rho_{DM}^2$ as for annihilation) there is limited sensitivity
to the choice of the profile.  For example, another standard profile,
the NFW profile
\beq
\rho_{DM}(r) = \rho_0 \left(\frac{r_s}{r}\right) ^\gamma \left(1+\left (\frac{r}{r_s}\right) ^\alpha \right) ^{(\gamma -\beta)/\alpha}
\eeq
with parameter choices $\{\alpha,\beta,\gamma\} =\{1, 3, 1\}$, $r_s =
20 \,\kpc$, and $\rho_0 = 0.26\gev/{\rm cm}^3$ \cite{nfw}, increases
the neutrino flux above that predicted using the Einasto profile by
about 10 percent.

\subsection{Neutrino bounds from Super-Kamiokande}

Dark matter decaying in our galaxy produces mostly neutrinos and
antineutrinos with energy equal to $m_\lambda/2$.  This leads to a
flux of cosmic neutrinos concentrated in the direction toward the
galactic center.  The interaction cross-sections of dark matter with
nuclei in our model are negligible, leading to correspondingly
negligible dark matter capture rates in the Earth and the Sun.
Therefore only bounds on neutrino flux from the galactic center are
constraining.  To date, the best bounds on such a flux are those from
the Super-Kamiokande collaboration \cite{superk,Desai:2007ra}.

The flux of neutrinos coming from dark matter decay is given by
\beq
\frac{d\Phi_\nu}{dE_\nu d\Omega} (m_\lambda) =  \frac{1}{4\pi} \frac{\Gamma_\lambda}{m_\lambda}\frac{dN}{dE_\nu} \int ds\,\rho (s),
\eeq
where $dN/dE_\nu$ is the spectrum of neutrinos generated in the dark
matter decay, and the integral of the dark matter energy density is
taken along a line of sight.  It is conventional to define the
dimensionless quantity
\beq
\frac{dJ}{d\Omega} =\int\frac{ds\,\rho (s)}{r_\sun \rho_\sun}\, ,
\eeq 
normalizing the line of sight density integral to our local position
and density.  The flux in a cone of half-angle $\psi$ around the
galactic center is then
\beq
\label{e.gcf}
\frac{d\Phi_\nu}{dE_\nu} (m_\lambda; \psi)= \frac{1}{4\pi} \frac{\Gamma_\lambda}{m_\lambda}\frac{dN}{dE_\nu}  r_\sun \rho_\sun \times\Delta J(\psi),
\eeq
with
\beq
\Delta J(\psi)=\frac{2\pi}{ r_\sun \rho_\sun } \int^1 _{\cos\psi} d\cos\theta \int_0 ^ {r_0} dx\;\rho(\sqrt{r_\sun^ 2+x ^ 2-2r_\sun x\cos\theta}) .
\eeq
Notice that \eref{gcf} is the total neutrino flux, summed over all
three flavors, as oscillation over galactic scales will equally
populate all three flavors. 
Moreover, there is an equal in magnitude anti-neutrino flux.

Having fixed the halo profile in \eref{ep}, $\Delta J$ is a function
of the half-cone angle which can be computed numerically.  Neutrinos
produced in a decay travel to the Earth without experiencing further
interactions.  The normalization of the neutrino flux at the
Earth depends on the dark matter lifetime and mass via the ratio
$\Gamma_{2b}/m_\lambda$, and this is the quantity that will be
constrained by Super-Kamiokande.

The published Super-Kamiokande bounds do not apply directly directly
to the neutrino flux but rather to the upward-going muon flux.  These
muons are produced by charged current interactions of the cosmic
neutrinos in the rock below the detector.  In order to translate the
neutrino and antineutrino fluxes into a muon flux we need to compute
the conversion functions $C_{\nu \to \mu^-}(E_\mu; E_\nu)$ and
$C_{\bar \nu \to \mu^+}(E_\mu; E_\nu)$,
\beq
\label{e.cfDef}
{d \Phi_{\mu^-} \over d E_\mu} =    \int dE_\nu {d C_{\nu \to \mu^-} \over d E_\mu} {d \Phi_\nu \over d E_\nu} \:,
\qquad 
{d \Phi_{\mu^+} \over d E_\mu}  =    \int dE_\nu  {d C_{\bar \nu \to \mu^+} \over d E_\mu } {d \Phi_{\bar \nu} \over d E_\nu}\:.
\eeq
which encapsulate the probability that a neutrino will interact in
rock to produce a muon of energy $E_\mu$ which reaches the detector.
The conversion functions depend on the the muon range in rock
$R_\mu(E_\mu)$, the proton and neutron number densities in rock
$n_{N}$, $N = n,p$, and the neutrino inelastic scattering cross
section on nucleons $\sigma_{\nu N}$:
\beq 
\label{e.dupa1}
{d C_{\nu \to \mu^-} \over d E_\mu} =  R_\mu(E_\mu) 
\left [n_p {d\sigma_{\nu p} \over dE_\mu}   +  n_n {d \sigma_{\nu n} \over dE_\mu} \right ]\:,
\qquad
{d C_{\bar \nu \to \mu^+} \over d E_\mu} =  R_\mu(E_\mu) 
\left [n_p {d\sigma_{\bar \nu p} \over dE_\mu}   +  n_n {d \sigma_{\bar \nu n} \over dE_\mu} \right ] .
\eeq 
%
To compute the neutrino-nucleon scattering cross sections, we use the
tree-level partonic cross-section integrated over the CTEQ5 parton
distribution functions.  We perform our computations using the
fiducial material ``standard rock'', for which the nucleon densities
are $n_p = n_n \approx 6.1 \times 10^{-18} \gev^3$.

The last piece we need to evaluate \eref{dupa1} is the muon range
$R_\mu(E_\mu)$, the distance a muon with the initial energy $E_\mu$
can travel before its energy is radiated away so as to fall below the
Super-Kamiokande threshold $E_0 = 1.6$ GeV.  The muon range can be
modeled analytically using the ``continuous slowing down
approximation'', with parameters found in \cite{Groom:2001kq}.  We
approximate the muon range $R_\mu(E_\mu)$ as
\beq
\label{eq:range}
R_\mu(E_\mu ) \approx 10 ^ 5 \cm \; \ln \left(\frac{E_\mu+\epsilon}{\epsilon}\right),
\eeq
Here $\epsilon$ is the critical energy, where ionization and radiation
loss are equal; we take $\epsilon$ to be $693 \gev$ in standard rock
\cite{Groom:2001kq} (we assume that  cross sections in standard 
rock are comparable to cross sections in the rock below Super-Kamiokande).  
The continuous slowing down approximation
becomes poor at high energies, where stochastic radiative processes
dominate the energy loss.  More accurate Monte Carlo modeling of muon
propagation at high energies reduces the flux of upward-going
neutrino-induced muons at neutrino detectors for a given incident
neutrino flux relative to the flux obtained using the continuous
slowing down approximation \cite{Lipari:1991ut}. Comparing the
effective muon range given by \er{range} to the effective muon range
computed in \cite{Lipari:1991ut} for muons with energy $>\tev$, we
find our muon ranges are overestimated by a factor of less than order
unity over most of the energy range.  The bounds we obtain from
Super-Kamiokande's reported upward through-going muon counts are 
therefore  conservative.  

Finally, to obtain the total muon flux in the Super-Kamiokande
detector we have to integrate over all muon energies between $E_0$ and
$m_\lambda/2$.  At the end of the day the muon flux in
Super-Kamiokande is given by
\beq
\label{e.mfsk} 
\Phi_{\mu^\pm}(\psi) =  {1 \over 3} {\Gamma_{2b} \over m_\lambda}  {r_{\sun} \rho_{\sun} \over 4 \pi}  \Delta J_{ls}(\psi)
\int_{E_0}^{m_\lambda/2} dE_\mu \left [{d C_{\nu \to \mu^-} \over d E_\mu} +  {d C_{\bar \nu \to \mu^+} \over dE_\mu} \right ] 
\eeq
The factor $1/3$ arises because $2/3$ of the produced neutrinos
oscillate into electron and tau neutrinos to which Super-Kamiokande is
far less sensitive.

\begin{figure}[tb]
\bc 
\includegraphics[width=0.5\textwidth]{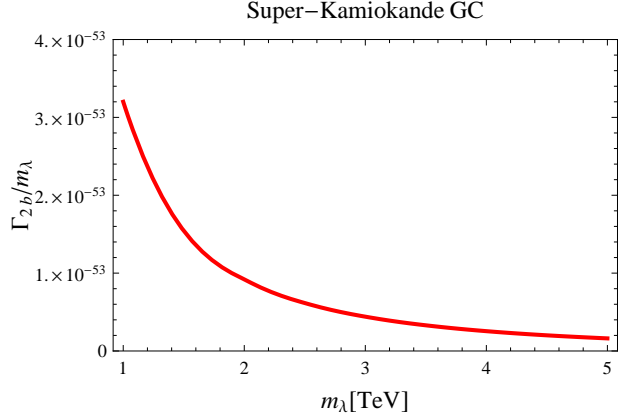}
\ec 
\caption{The maximal $\Gamma_{2b}/m_\lambda$ as a function of $m_\lambda$ allowed by the Super-Kamiokande bounds.} 
\label{f.superKGamma}
\end{figure}

The muon flux induced by neutrinos from the galactic center predicted
by our model is dominated by the monoenergetic neutrinos from the
dominant two-body decay.  At higher dark matter masses, $m_\lambda
\sim 5\tev$, the additional neutrinos from the sub-dominant three-body
modes begin to make a noticeable contribution to the total muon flux.
We computed the total muon flux predicted from our model and
confronted it with the Super-Kamiokande bounds from \cite{superk}.
The most stringent bounds are obtained for cones of half-angle $30$
degrees.  In \fref{superKGamma} the resulting bound on
$\Gamma_{2b}/m_\lambda$ is plotted as a function of $m_\lambda$.  The
dark matter mass is varied in the range $1-5$ TeV, as motivated by the
estimate of thermal abundance in the previous section.  For $m_\lambda
= 1$ TeV the bound is $\Gamma_{2b}/m_\lambda \simlt 3.2 \times
10^{-53}$, corresponding to the lifetime $\tau \approx 9.9\times
10^{24}$ seconds.  For $m_\lambda = 5$ TeV the bound is
$\Gamma_{2b}/m_\lambda \simlt 1.6 \times 10^{-54}$, corresponding to
the lifetime $\sim 1.9 \times 10^{25}$ seconds.

The sub-dominant three-body decays lead to additional visible signals,
in particular, to antiprotons, hard leptons (which we take here as
$e^\pm$), and photons.  In the following subsections we investigate
the predicted antiproton, positron and gamma ray fluxes arising from
the three-body decay modes to determine whether these subleading
decays impose more stringent bounds on the lifetime of our dark matter
particle than the two-body decays.

\subsection{Antiproton bounds from PAMELA}

An irreducible source of antiprotons in our model are the decays of
the $W$, $Z$, and Higgs bosons which are produced by the three-body
decays of our dark matter particle.  On average, these channels lead to one
energetic antiproton per decay.  The branching ratio for three-body
decays varies from 4 percent of the total width for $m_\lambda = 1$
TeV to as much as 55 percent for $m_\lambda = 5$ TeV.  Thus we expect
that the antiproton bounds on the dark matter lifetime can become
quite stringent in some regions of the parameter space.

The production rate of antiprotons per unit energy due to the
three-body decays is given by
\beq
\label{eq:qdef}
Q_{\bar p}(E,\vec r) = 2 {\Gamma_{3b} \over m_\lambda} \rho_{DM}(\vec r)  
\left ({dN_{Z \to \bar p} \over d E}  + {dN_{h \to \bar p} \over d E} + 2 {dN_{W \to \bar p} \over d E}  \right ).  
\eeq 
We determined the antiproton spectra $dN_{\cdot \to p}/d E$ by first generating $W$,
$Z$, and Higgs decays at rest using Pythia 6.420, and then
boosting the momenta of antiprotons to the galactic rest frame.  The
energy distribution of the parent gauge and Higgs bosons is given by
the differential decay width, which is the same for all three
particles in the limit $m_\lambda\gg m_{h,W,Z}$:
\beq
\frac{1}{\Gamma_{3b}} \frac{d\Gamma_{3b}}{d E} =\frac{32}{m_\lambda ^ 4} E (m_\lambda-2 E) (m_\lambda -E).
\eeq
Throughout, the Higgs is assumed to have standard model properties and the mass equal 115 GeV.

In order to translate the production rate into the flux of antiprotons
at the Earth it is necessary to model the propagation of antiprotons
in our galaxy \cite{DMS}.  The antiproton number density per unit
energy $N_{\bar p}$ satisfies a diffusion equation
\beq
\label{e.appe} 
K(E)\nabla^2 N_{\bar p}(E,\vec r) - \pa_z (V_c {\rm sgn}(z) N_{\bar p}(E,\vec r)) - 2 h \delta(z)  \Gamma_{ann}(E) N_{\bar p}(E,\vec r) 
+ Q_{\bar p}(E,\vec r) = 0
\eeq 
in which the production rate \er{qdef} appears as the source term.
Energy loss for protons is negligible and the corresponding term in
the diffusion equation is dropped.  The diffusion coefficient $K(E)$
is parameterized as $K(E) = K_0 (E/\gev)^\delta$.  The parameters
$K_0$ and $\delta$ are not known from first principles; instead, they
are fitted, along with the magnitude of the convective wind $V_c$ and
the size of the diffusion zone, so as to correctly describe the
observed flux of cosmic rays.  The third term describes the
disappearance of antiprotons due to annihilations with protons in the
galactic disk.  The half-height of the galactic disk is set by $h =
0.1 \;\kpc$, and the annihilation width is given by the number
densities of hydrogen and helium multiplied by the proton-antiproton
annihilation cross-section, $\Gamma_{ann}(E) = (n_H + 4^{2/3} n_{He})
\sigma_{p\bar p}(E)$.  Expressions for $n_H, n_{He}$ and the
proton-antiproton annihilation cross-section $\sigma_{p\bar p}(E)$
can be found in \cite{Hisano:2005ec}.  The diffusion equation
\eref{appe} is solved subject to the boundary condition that the
antiproton density $N_{\bar p}$ vanish at the surface of the
diffusion region, taken to be a cylinder of radius $R_{max}=20 \;\kpc$ and half-height $L_{max}$.

In the following we focus on the MED propagation model
\cite{Donato:2003xg}, in which case the diffusion equation parameters
take the following values:
\beq
K(E) = 3.5 \times 10^{-10} \kpc^2/\mathrm{s} ; \:\:\: \delta = 0.7 ; \:\:\: V_c =1.2 \times 10^6 \cm/\mathrm{s} ; 
   \:\:\: L_{max} = 4 \;\kpc . 
\eeq
%
%
%
Once we know $N_{\bar p}$, the flux of  antiprotons
observed at the Earth is given by ${d \Phi_{\bar p} \over d E} = {v
\over 4 \pi} N_{\bar p}(E,\vec r_{\sun})$; in the interesting energy range the antiprotons are highly relativistic and $v \approx 1$. 
    
The propagation equation can be solved semi-analytically
\cite{BBDMST}.  Using that solution, the antiproton flux in our model
is given by 
\beq
{d \Phi_{\bar p} \over d E} = {1 \over 2 \pi}  {\Gamma_{3b} \over m_\lambda}
\left ({dN_{Z \to \bar p} \over d E}  + {dN_{h \to \bar p} \over d E} + 2 {dN_{W \to \bar p} \over d E}  \right ) 
\rho_{DM}(r_{\sun}) 
R(E) .
\label{e.pbarf} 
\eeq 
The propagation effects are encoded in the function $R(E)$ which can
be computed numerically \cite{BBDMST,Donato:2003xg}.

\begin{figure}[tb]
\bc 
\includegraphics[width=0.3\textwidth]{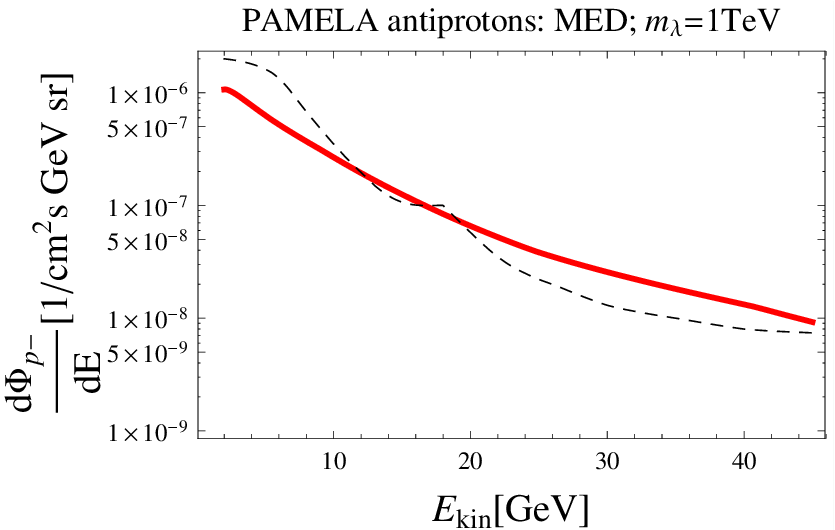}
\includegraphics[width=0.3\textwidth]{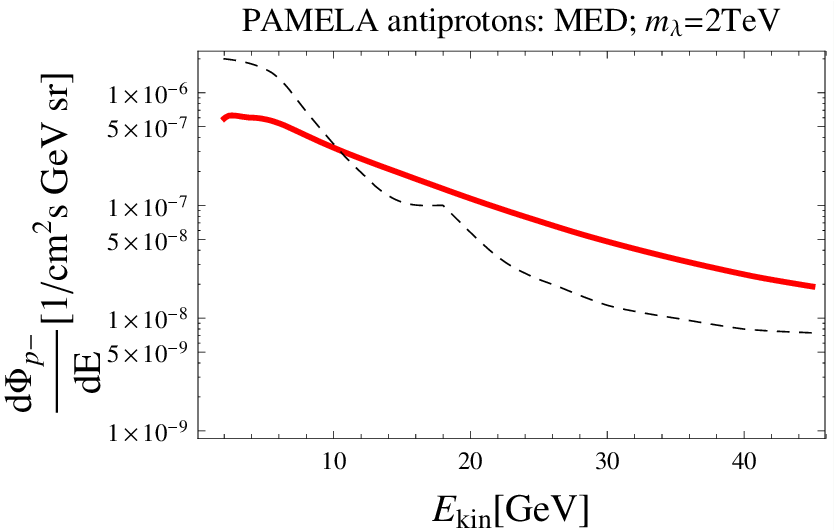}
\includegraphics[width=0.3\textwidth]{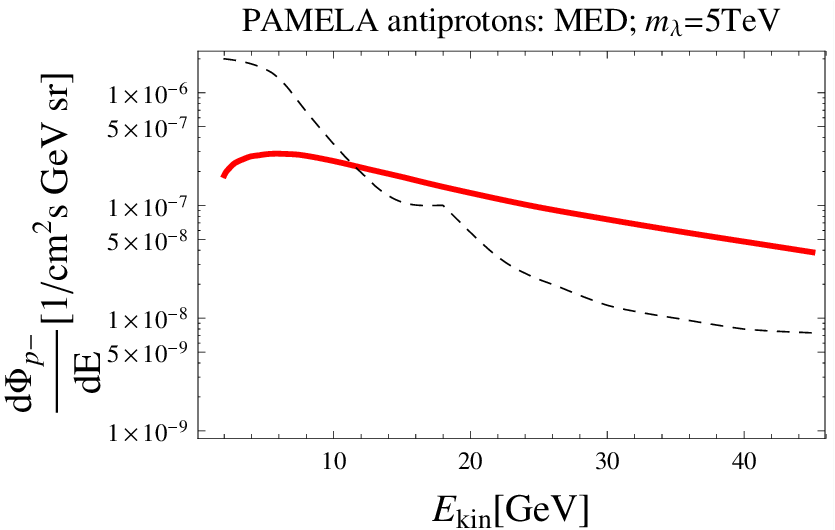}
\ec 
\caption{Antiproton flux from the three-body decay of the gauginos (solid red) compared to the flux measured by PAMELA (dashed) for dark matter masses $m_\lambda =1$, 2, and 5 TeV. In each case the glueballino decay rate is set to the maximal value allowed by the Super-Kamiokande bounds.}
\label{f.pbar3b_flux}
\end{figure}

With this solution in hand we are ready to estimate the antiproton
constraints on the lifetime of our dark matter particle.  We pick
three representative values for the dark matter mass: $m_\lambda =
1,2,5$ TeV.  For the sake of comparison with the neutrino bounds, for
each mass we choose the maximum $\Gamma_{2b}/m_\lambda$ allowed by
Super-Kamiokande as plotted in \fref{superKGamma}. Given $\Gamma_{2b}$
and $m_\lambda$, the partial width for three-body decays $\Gamma_{3b}$
is completely determined by equation \er{2to3}.  The mass fixes the
spectra $dN/dE$, while the partial width fixes the normalization of
the antiproton flux in \eref{pbarf}.  Our results are plotted in
\fref{pbar3b_flux} and compared to the antiproton flux measured by
PAMELA.  It is clear that antiprotons, despite being produced by
subdominant decay channels, can put important constraints on the dark
matter lifetime.  For $m_\lambda = 5$ TeV, the predicted antiproton
flux is 5 times larger than observed, which would imply the
stringent bound $\Gamma_{2b}/m_\lambda \simlt 3 \times 10^{-55}$
corresponding to the lifetime $\sim 10^{26}$ sec.  For $m_\lambda = 1$
TeV, the predicted antiproton flux is only marginally larger than
observed, and is slightly more constraining than the Super-Kamiokande
bound.  One should however keep in mind that the propagation of
antiprotons in our galaxy suffers from huge uncertainties, and the
predicted antiproton flux changes dramatically with different
assumptions about the propagation model.  For example, choosing the
MIN parameters for the propagation model \cite{Donato:2003xg},
\beq
K_0 = 5.1 \times 10^{-11} \kpc^2/\mathrm{s};  \:\:\:  \delta = 0.85;  \:\:\: V_c = 1.35 \times 10^6 \cm/\mathrm{s} ;
\:\:\: L_{max} = 1 \;\kpc ,
\eeq
reduces the antiproton flux by a factor of five.  This would make the
antiproton bounds comparable to the neutrino ones for $m_\lambda = 5$
TeV, and irrelevant for $m_\lambda = 1$ TeV.  (It is of course also 
possible to choose propagation parameters which magnify the antiproton flux.)
Thus, the antiproton bounds should be taken with a grain of salt.  
In the following, we continue using the more robust Super-Kamiokande 
constraints as the reference point for estimating the indirect signals 
of dark matter.

\subsection{Other constraints: photons, positrons}

Our decaying dark matter particle contributes also to the
cosmic flux of positrons and photons.  Hard positrons are produced
together with $W$ bosons in three-body decays and, to a lesser extent, by
subsequent decays of $W$, $Z$, and Higgs.  Gamma ray photons are
produced via decays of $W$, $Z$, and Higgs, from final state radiation
off charged particles, and through inverse Compton scattering of the
hard positrons and electrons on starlight.  Below we compute the
fluxes of hard positrons and photons from the three-body decay modes and
argue that the corresponding bounds are far weaker than those from
neutrinos and antiprotons.  Therefore, the PAMELA positron
measurements and the gamma ray measurements from FERMI and HESS do not
impose new constraints on the lifetime of our dark matter particle.

The positron flux at the Earth due to the positrons from three-body 
decays is given by \cite{Delahaye:2007fr}
\beq 
{d \Phi_{e^+} \over d E} = 
{\Gamma_{3b} \over m_\lambda} {1 \over 2  \pi b(E)}  \rho_{DM}(r_{\sun}) 
\int_E^{m_\lambda/2} d E' {dN_{e^+} \over d E'}  I(E;E') 
\label{e.PositronFlux}
\eeq   
Here, the energy loss coefficient $b(E)$ gives the rate of energy loss
due to scattering off of cosmic photons and magnetic fields, and is
approximately given by $b (E) =\tau_E E^2/\gev $ where $\tau_E =
10^{16}$s. The function $I (E; E') $, sometimes called the halo
function, results from semi-analytically solving the diffusion
equation for positrons \cite{Delahaye:2007fr}.  It is analogous to the
function $R (E) $ for antiprotons, and similarly depends on the
choice of propagation model.  Finally, $dN_{e^+}/d E$ is the positron
spectrum from three-body decays, given by
\beq
\frac{1}{\Gamma_{3b}} \frac{d\Gamma_{3b}}{d E} =\frac{32}{m_\lambda ^ 4} E^ 2 \left(m_\lambda-\frac{2}{3} E\right) .
\eeq
%
%

To obtain the positron fraction one has to divide the flux in
\eref{PositronFlux} by the total flux of electrons and positrons, for
which we take
\beq
\frac{d\Phi_{e ^\pm}}{d E} =\frac{0.012}{E^3}\; 1/(\cm^2 \sec \gev)
\eeq
as measured by FERMI \cite{Abdo:2009zk}.  As an example, in
\fref{eplus3b_flux} we plot the contribution to the positron fraction
from the three-body decays for $m_\lambda = 2$ TeV, using the MED
propagation parameters.  We took the decay rate
$\Gamma_{3b}/m_\lambda$ to be the maximum one allowed by the neutrino
bounds from Super-Kamiokande.  Recall that for these parameters
antiprotons are overproduced by a factor of 2--3.  On the other hand,
the positron fraction for the same propagation parameters is a factor
of 2--4 below that measured by PAMELA.  We conclude that positrons do
not impose any new constraints on the lifetime of our dark matter
particle.  At the same time, we cannot explain the anomalous positron
fraction via the positrons from three-body decays; another primary
source of positrons is needed to explain the PAMELA positron
measurement.  The new source could be of purely astrophysical origin
\cite{Grasso:2009ma}, but it could be glueball (or meson) decays in
our model.
 
\begin{figure}[tb]
\bc 
\includegraphics[width=0.3\textwidth]{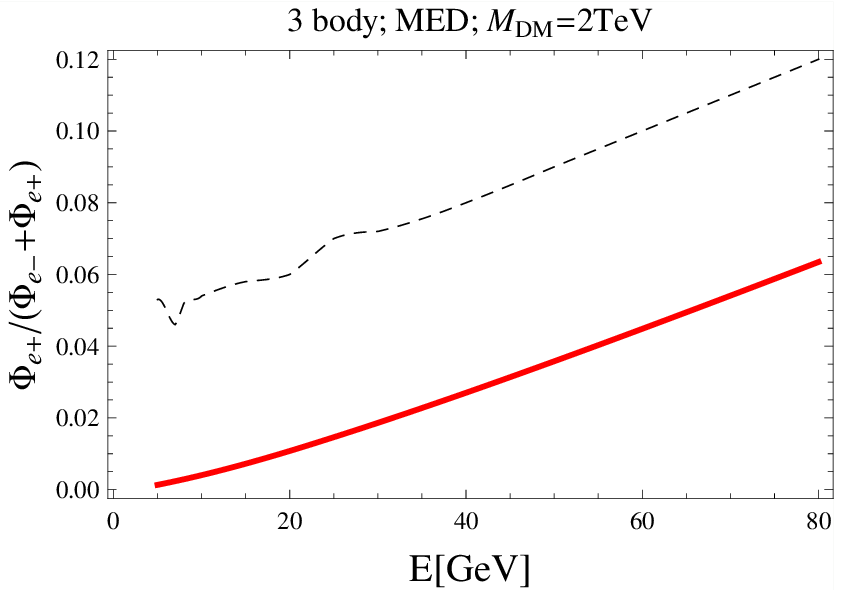}
\includegraphics[width=0.35\textwidth]{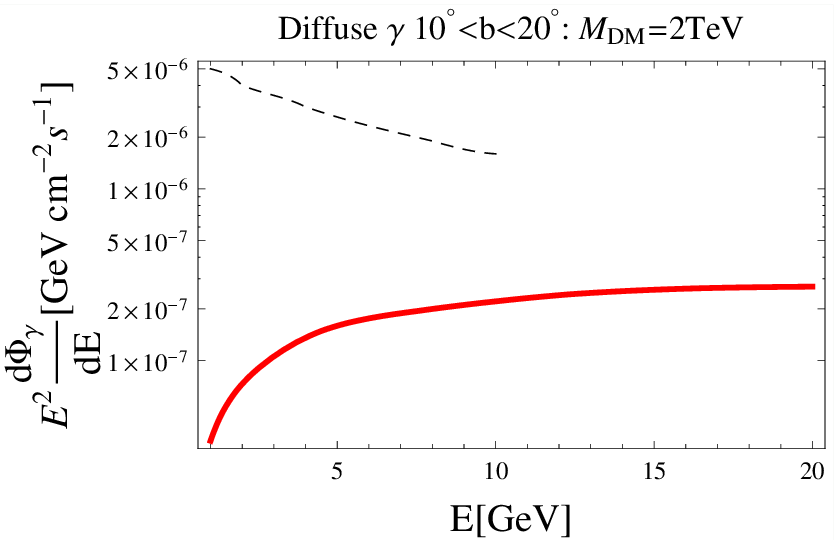}
\ec 
\caption{Contributions from three-body decays to positron and gamma ray fluxes for  $m_\lambda = 2$ TeV. 
Left: the positron fraction (solid red) compared to the PAMELA measurement (dashed). 
Right: the diffuse gamma ray flux  $10^\circ < b <20^\circ$, $0 \leq l \leq 360^\circ$ (solid red) compared to the FERMI measurement (dashed).}
\label{f.eplus3b_flux}
\end{figure}

Similarly, the contribution to the gamma ray spectrum from the
three-body decays is well below the current observational bounds.  As
an example, we calculate the contribution to the diffuse gamma ray
spectrum at intermediate galactic latitudes.  Recently, FERMI
published the gamma ray spectrum between 0.1 and 10 GeV from between
10 and 20 degrees galactic latitude \cite{Porter:2009sg}.  The
contribution of three-body decays to that spectrum is given by
\beq
{d \Phi_{\gamma} \over d E } =  2 {\Gamma_{3b} \over m_\lambda} {d N_{\gamma} \over d E} I_{d}(10^\circ,20^\circ)
\qquad 
I_{d}(\psi_1,\psi_2) = {1 \over 4 \pi} \int_0^{2\pi} d \phi\int_0^{\infty} dl \int_{\cos\psi_1}^{\cos \psi_2} d \cos \theta \rho_{DM}(r) 
\eeq   
where $r = \sqrt{r_{\sun}^2 + l^2 - 2 r l \cos \theta \cos \phi }$.
Using Pythia, we determined the photon spectrum ${d N_{\gamma}/d E}$
from $W$, $Z$, and Higgs decays, where the parent bosons originate
from the three-body decays of dark matter. The results for $m_\lambda
= 2$ TeV are plotted in \fref{eplus3b_flux}.  It is clear that the
contribution of the three-body decays is more than an order of
magnitude smaller than the flux measured by FERMI.  Similarly, the
contribution to the gamma ray flux from the galactic center is far
below the flux measured by FERMI and HESS \cite{HESS}.  Thus, the strongest
constraints on the dark matter lifetime in our model are the galactic
center neutrino measurement by Super-Kamiokande and the cosmic
antiproton flux measurement by PAMELA.

\section{Discovery at Neutrino Telescopes}

Our model predicts a monochromatic flux of high-energy ($E_\nu 
\sim \tev$) neutrinos and antineutrinos from the galactic center,   
\beq  
\label{e.mnf} 
{d \Phi_{\nu} \over d E }(\psi) = {d \Phi_{\bar \nu} \over d E }(\psi) =  
{\Gamma_{2b} \over m_\lambda} \delta (E - m_\lambda/2)  {r_{\sun} \rho_{\sun}\over 4 \pi} \Delta J_{ls}(\psi),
\eeq 
where the line-of-sight integral over the dark matter density profile
$\Delta J_{ls}(\psi)$ is given in \eref{gcf}.  The normalization of
the flux is set by the dark matter lifetime, $\Gamma_{2b}$.  The flux
is equally populated by all three species of neutrinos; the flux of
muon neutrinos to which experiments are most sensitive is one third of
\eref{mnf}.  The Super-Kamiokande results place constraints on the
dark matter lifetime $\Gamma_{2b}/m_\lambda$, as shown in figure
\fref{superKGamma}.  The dark matter lifetime is also constrained by
the antiprotons produced in subdominant three-body decay modes.
While the limits on cosmic antiprotons can be comparable to or more
stringent than the bounds from Super-Kamiokande, depending on the
choice of model for charged particle propagation in the galaxy, to
discuss signals at neutrino telescopes we will continue to use the
limits on the dark matter lifetime set by Super-Kamiokande.  In this
section we discuss the sensitivity of the ANTARES and IceCube neutrino
telescopes to the primary neutrino signal.  Both experiments have the
potential to discover this signal.

\subsection{ANTARES}

ANTARES is a high-energy neutrino telescope located in the
Mediterranean sea.  The flux of muon neutrinos
from the galactic center produces upward-going muons via charged
current interactions in the rock and water below the detector, which ANTARES
observes via the Cherenkov light produced in sea water by the muons.
The effective detector area $A_{eff}(E)$ for muon neutrinos can be
found e.g. in \cite{Brown:2009hu} and at 1 TeV is given approximately
by $50 \cm^2$.  The energy resolution near 1 TeV is expected to be a
factor of 2--3 \cite{Ernenwein:2005ny}, while the angular resolution is better
than a degree and limited mostly by the kinematics of
the charged current interactions \cite{Brown:2009hu}.

The number of signal counts per second from the cone of half-angle 
$\psi$ around the galactic center is given by 
\beq  
{d N_{signal} \over d t}  =    {2 \over 3}  A_{eff}(m_\lambda/2) {\Gamma_{2b} \over m_\lambda}  {r_{\sun} \rho_{\sun} \over 4 \pi} \Delta J_{ls}(\psi), 
\eeq 
and the neutrino energy is $m_{\lambda}/2$. 

The irreducible background for the dark matter decay signal comes from
atmospheric muon neutrinos.  The combined flux of atmospheric muon
neutrinos and anti-neutrinos at $1\tev $ is, averaged over angle
\cite{HKKMS},
\beq
\left.\frac{d\Phi_{atm}}{dEd\Omega}\right|_{1\tev} = 2.7\times 10 ^{-11} (\gev \mathrm{cm}^ 2\mathrm{s\; sr}) ^{-1}.
\eeq
The atmospheric neutrino spectrum falls as $E^{-3}$ in the energy
range of interest, $100\gev <E\lsim 10\tev $.  In the energy range
$500\gev <E \lsim 5\tev $ where we expect hard primary neutrinos from
decaying dark matter, neutrino telescopes have poor energy
resolution. To differentiate signal from background we therefore rely
on first, absolute rate, and second, angular variation in the signal.
The neutrinos coming from dark matter decay are concentrated at the
galactic center, while the spatial variation of the atmospheric
neutrino fluxes is well-known and is at most a factor of $2.5$ between
zenith and horizon \cite{HKKMS}.  The number of counts per
second from atmospheric neutrinos in a given energy bin $E_{min},\,
E_{max}$ is given by
\beq  
\frac{dN_{atm}}{dt d\Omega} =  \int_{E_{min}}^{E_{max}} dE A_{eff}(E)  {d\Phi_{atm} \over d E d \Omega }
\eeq  
We approximate the energy resolution of ANTARES by choosing an energy
bin $m_{\lambda}/6 < E < 3 m_{\lambda}/2$, corresponding to a factor
of three in energy resolution centered on the energy of the primary
neutrino signal.  In \fref{antares} we plot the number of signal and
background counts per year in this bin for cones around the galactic
center as a function of the cone half-angle.  We show plots for
$m_\lambda = 1,\,2,$ and $5$ TeV.  To generate these plots, we assume 
that the detector effective area $A_{eff}$ already incorporates 
factors accounting for detection efficiency and the different cross 
sections of neutrinos and anti-neutrinos.  We use the atmospheric 
$\nu_\mu$ + $\nu_{\bar \mu}$ flux $d\Phi_{atm}/d E d \Omega$ computed 
in \cite{HKKMS} and averaged over the zenith angle.  For the signal, 
we show expected counts for the maximal decay rate allowed by the
Super-Kamiokande bounds.  For that signal rate, the five-degree cone
corresponds to on the order of 5 signal counts and a signal-to-background 
ratio in the range of 1--3 (for 1--5 TeV dark matter).  Thus, in the best 
case scenario, a clear signal can be detected already in one year of
ANTARES data.  If no signal is detected, ANTARES can significantly
improve the bounds on the dark matter lifetime in our scenario.  The
theoretical uncertainty in the atmospheric neutrino flux is of order
25 percent at 1 TeV \cite{HKKMS}.  From that we estimate that
one year of ANTARES data can improve the bounds on $\Gamma_{2b}$ by a
factor of 4 to 8, depending on the dark matter mass.

\begin{figure}
\bc 
\includegraphics[width=0.3\textwidth]{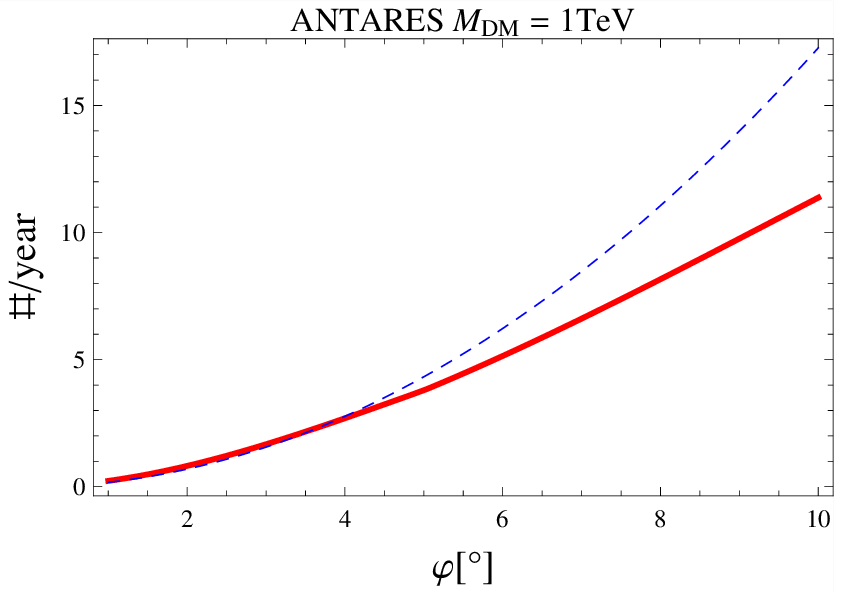}
\includegraphics[width=0.3\textwidth]{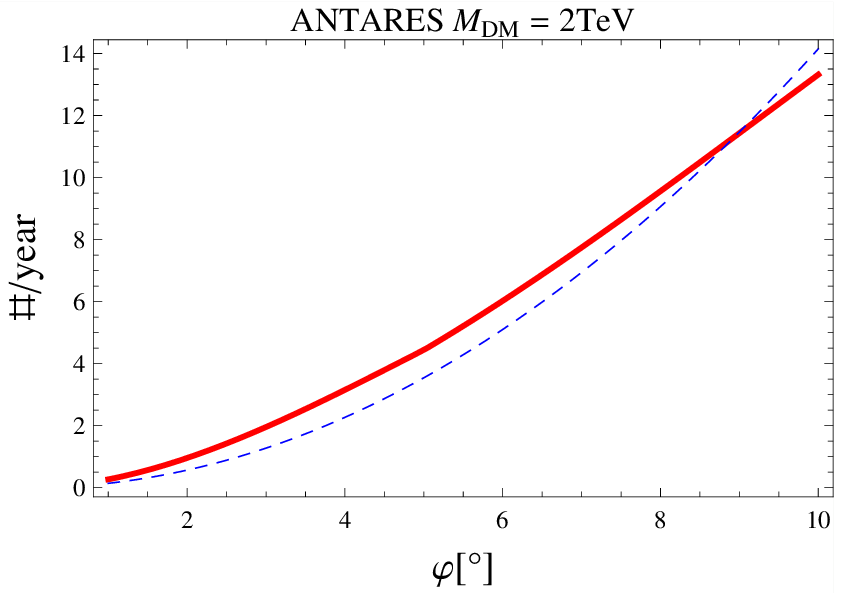}
\includegraphics[width=0.3\textwidth]{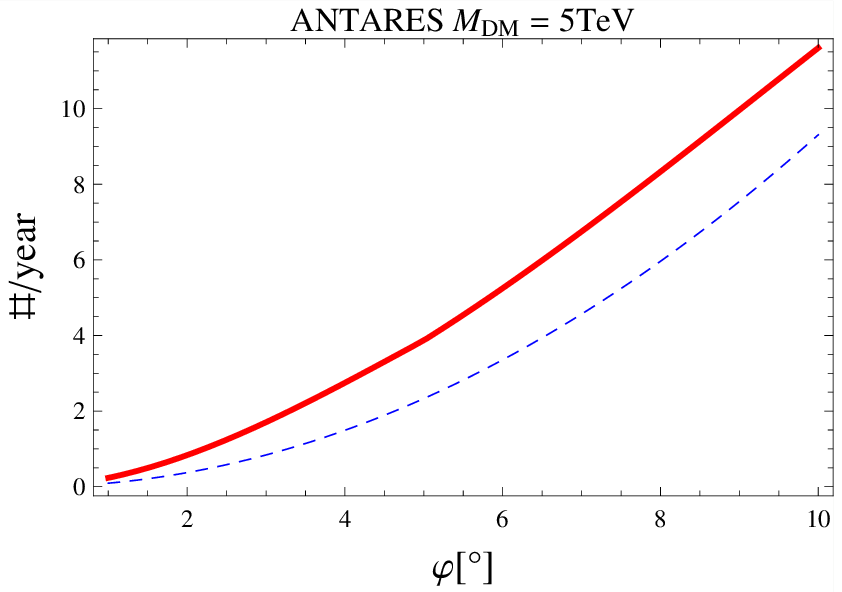}
\ec 
\caption{Neutrino counts per year in Antares from the signal (solid
red) and the background of atmospheric neutrinos (dashed blue) as a
function of the halfangle $\psi$ around the galactic center, for
$m_\lambda = $1, 2, and 5 TeV.  The normalization of the flux corresponds to
the maximal flux allowed by Super-Kamiokande for a given dark matter
mass.  }
\label{f.antares}
\end{figure}

\subsection{IceCube}

While IceCube, located at the South Pole, has better exposure to the
Northern Hemisphere, the extension DeepCore, planned for 2010, will
allow IceCube to observe the Southern Hemisphere sky
\cite{Resconi:2008fe}.  IceCube's new sensitivity to the galactic
center opens many exciting possibilities for dark matter studies \cite{buckley}.  DeepCore adds
additional instrumentation in the clearest and best-shielded portion
of the IceCube detector, and enables study of downward-going neutrinos
by using the remainder of the IceCube detector to veto cosmic muons.
Cosmic downward-going neutrino signals then consist of neutrino-initiated 
events with initial interaction vertices located within the fiducial volume of
DeepCore.

Muon neutrinos are primarily detected in DeepCore as they are in 
Super-Kamiokande and ANTARES, as a track from the muon produced in 
a charged-current interaction of the neutrino with matter.  
In addition to these {\em track-like events}, the neutral-current interactions
of neutrinos with all flavor, as well as the charged-current interactions of
electron and tau neutrinos, deposit energy in a localized region and
are reconstructed as {\em cascade-like} events. Cosmic neutrinos, and
in particular neutrinos from dark matter decay, populate all three 
flavors equally.  Usually at neutrino telescopes, the effective volume 
probed by muon neutrinos is orders of magnitude
larger than the effective volume seen by electron and tau neutrinos,
due to the large muon penetration depth.  However, when using DeepCore
to observe downward-going neutrinos, the volume available for muon
neutrino events is equal to the volume available for electron and tau
neutrino events, putting observable signal events from all three
neutrino flavors on an equal footing.  While angular resolution for
cascade events is less than angular resolution for tracks
\cite{Resconi:2008fe}, the high rate of cascade events predicted by
our model suggests another potential handle for discovery.

The fiducial volume of DeepCore is estimated to be around 20 megatons
\cite{Resconi:2008fe,Wiebusch:2009jf}, that is $V_F \approx 0.02\;
\mathrm{km}^3$.  Detection efficiencies for neutrino-initiated events
are estimated to be $20\%$--$30\%$ \cite{Resconi:2008fe}.  While the
angular resolution that can be achieved for DeepCore-initiating events
is still to be determined, a reasonable goal is 10 degrees for tracks,
possibly improving to 5 degrees, and 40 degrees for cascades
\cite{tyce}.  We will estimate the energy resolution to be $r_t = 0.5$
in $\log E$ for tracks and $r_c = 0.3$ in $\log E$ for cascades
\cite{tyce}.  The number of neutrino-initiated muons in DeepCore from 
a given incident neutrino flux is
\beq
\frac{dN_\mu}{dt d\Omega} = \int dE_\mu \int dE_\nu \,
                             \alpha_F V_F \sum_N n_N \left(\frac{d\Phi_\nu}{dE_\nu d\Omega} \, \frac{d\sigma_{\nu N}}{d E_{\mu}} 
                                + \frac{d\Phi_{\bar\nu}}{dE_{\nu} d\Omega} \,\frac{d\sigma_{\bar\nu N}}{d E_{\mu}}\right).
\eeq
The sum runs over nucleons, $N=n,\,p$. We take ice to have a density
$0.92 \;\mathrm{g}/\cm^3$, corresponding to the nucleon number
densities $n_n \approx 1.9 \times 10^{-18} \gev^3$, $n_p \approx 2.4
\times 10^{-18} \gev^3$.  We include a factor $\alpha_F = 0.25$ to
account for the detection efficiency.  The integral over the neutrino
energy is trivial for the signal; for the background we set the lower
integration limits at $E_{\nu } = e^{-r_{t,c}} m_\lambda/2$ for tracks
and cascades respectively.
 
In \fref{icecube} we plot the expected number of DeepCore-initiating 
track-like events originating within a cone of given half-angle around 
the galactic center.  For the maximal signal flux allowed by the Super-Kamiokande
constraints, one expects on the order of 25 (5) counts per year inside a 
cone of half-angle $30^\circ$ for a dark matter mass of 1 (5) TeV.  
The dominant background from muon atmospheric neutrinos, also plotted 
in \fref{icecube}, gives roughly the same number of track-like events 
as the signal.  However, it may be possible to reduce as much as 99 percent of the
atmospheric neutrino background by vetoing on the associated collinear
muon \cite{Schonert:2008is}.  If that proves to be the case, the 
background could be almost completely suppressed and
the discovery prospects for the signal are greatly enhanced.  The
sensitivity of IceCube would only be limited by statistics, and could
reach a lifetime of $10^{27}$ seconds after several years of
operation, two orders of magnitude better than the current bounds.

\begin{figure}
\bc 
\includegraphics[width=0.3\textwidth]{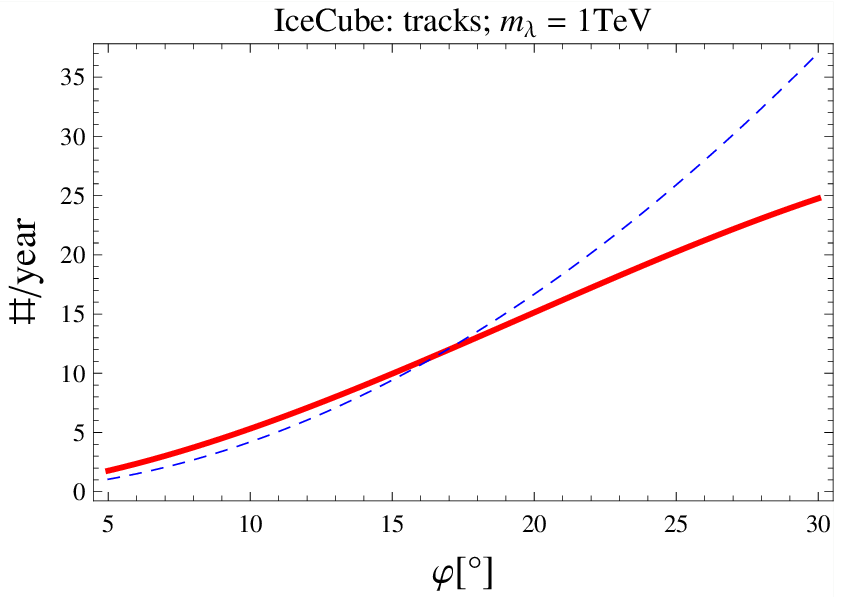}
\includegraphics[width=0.3\textwidth]{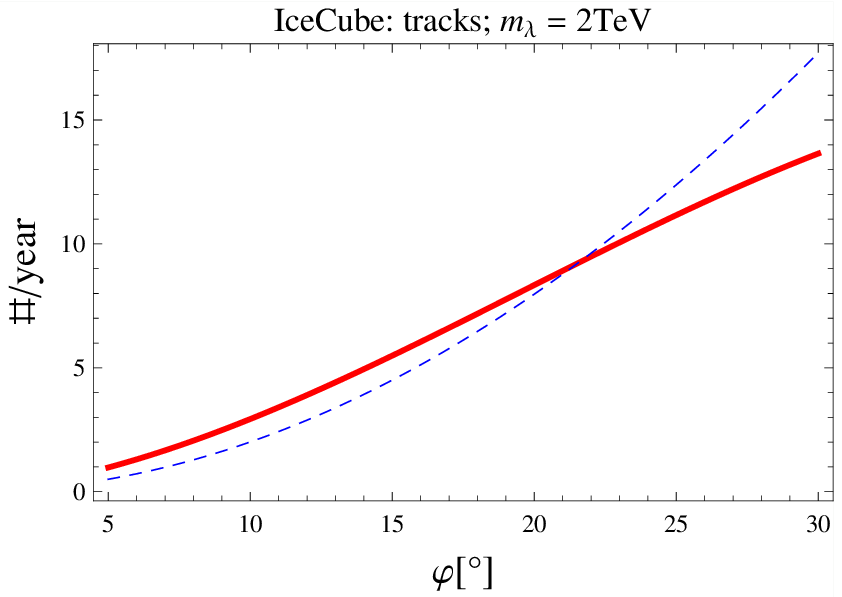}
\includegraphics[width=0.3\textwidth]{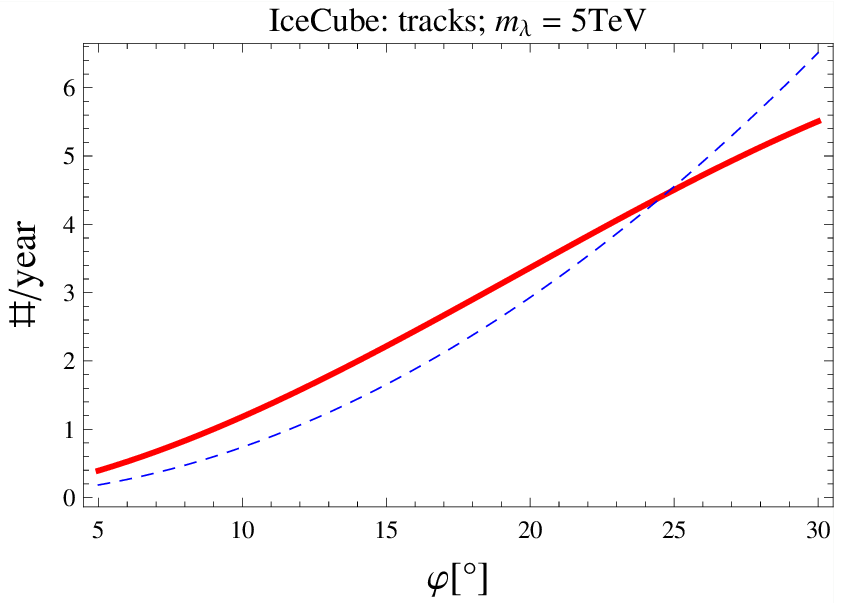}
\ec 
\caption{Track-like events per year in DeepCore due to the signal
(solid red) and the background of atmospheric muon neutrinos (dashed
blue) as a function of the half-angle $\psi$ around the galactic
center, for $m_\lambda = $ 1, 2, and 5 TeV.  The normalization of the flux
corresponds to the maximal flux allowed by Super-Kamiokande for a given
dark matter mass.  }
\label{f.icecube}
\end{figure}    

\begin{figure}
\bc 
\includegraphics[width=0.3\textwidth]{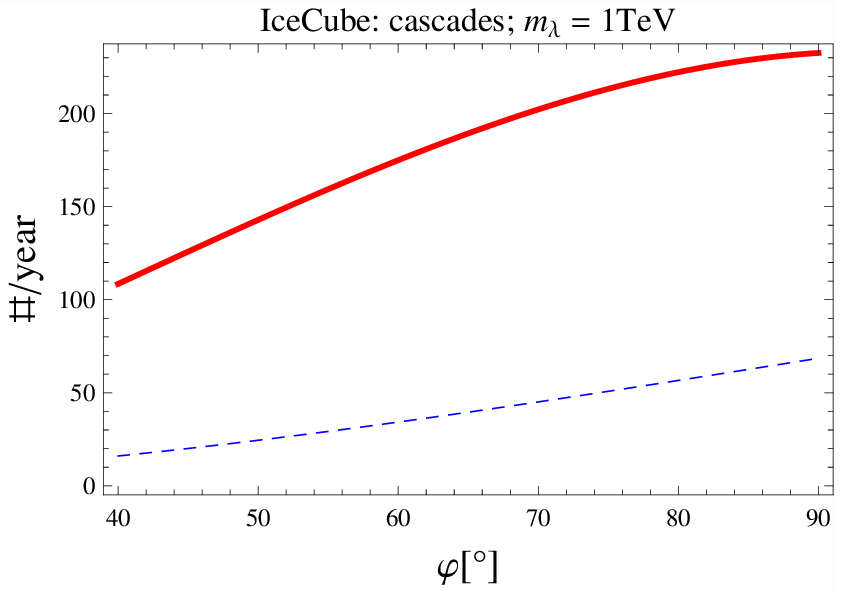}
\includegraphics[width=0.3\textwidth]{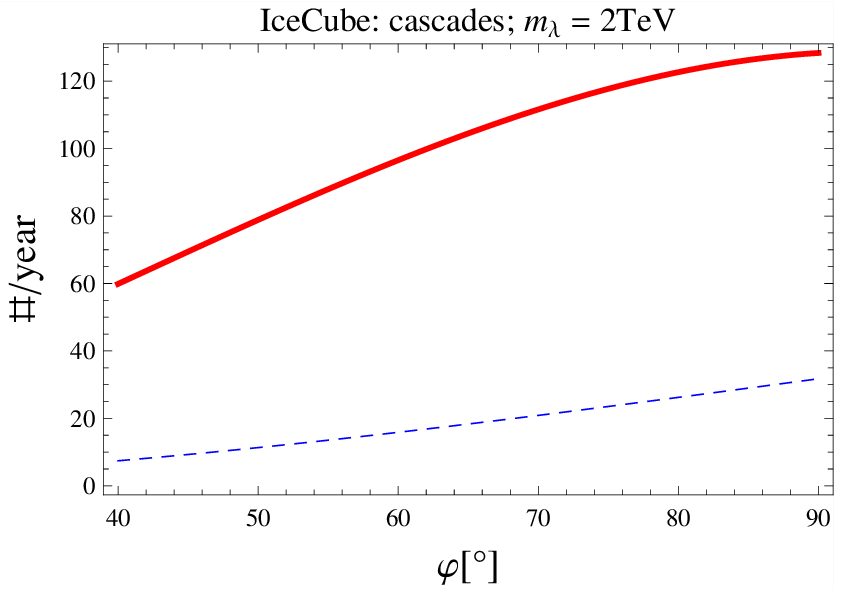}
\includegraphics[width=0.3\textwidth]{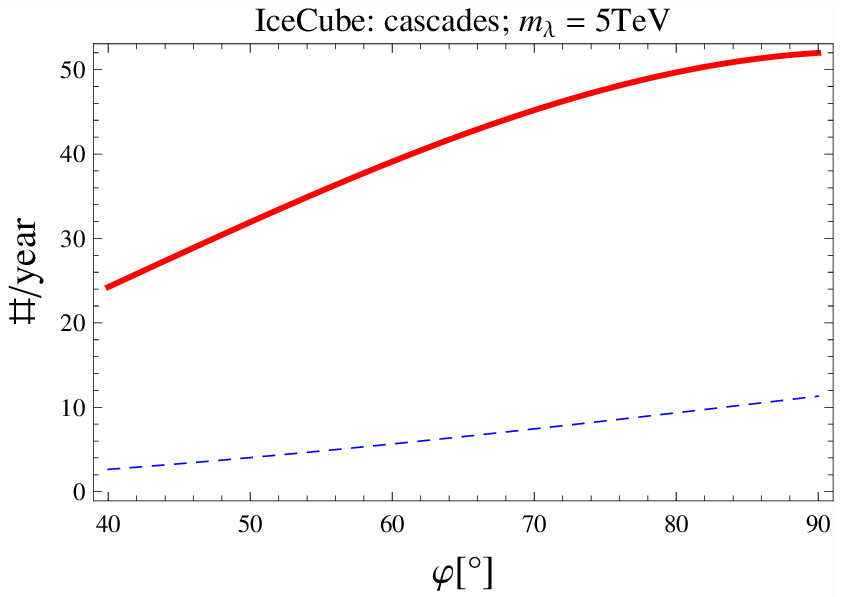}
\ec 
\caption{Cascade-like events per year in DeepCore due to the signal
(solid red) and the background of atmospheric neutrinos (dashed blue) as a
function of the halfangle $\psi$ around the galactic center, for
$m_\lambda = $1, 2, and 5 TeV.  The normalization of the flux corresponds to
the maximal flux allowed by Super-Kamiokande for a given dark matter
mass.  }
\label{f.cascades}
\end{figure}    

We also compute the number of cascade-like events initiated by the
signal and compare to events from the atmospheric neutrino background
As the atmospheric electron neutrino flux is approximately 5 percent 
of the atmospheric muon neutrino flux in the energy range of interest, 
the background is dominated by the neutral-current interactions of 
atmospheric muon neutrinos. Charged-current interactions of atmospheric 
electron neutrinos contribute approximately an eighth of the total background event 
rate.  For cascade events, rate alone can serve to distinguish signal 
from background, as the signal to background ratio can be much larger than one
for decay rates near the Super-Kamiokande bounds. In \fref{cascades} we plot
signal and background rates for cascade events for $m_\lambda = $ 1, 2, and 
5 TeV, for the maximal signal flux allowed by the Super-Kamiokande bounds.   
Again, it may be possible to nearly eliminate the background due to 
atmospheric muon neutrinos, which would reduce the background by nearly 
90 percent.  Thus cascade-like events, even more than track-like events, 
offer powerful handles on neutrino signals of this nature.  More importantly, 
observing an excess of both tracks and cascades would be an important confirmation of the galactic origin of the neutrino flux.

\section{Conclusions}
  
We presented a scenario where decay of hidden sector dark matter can give a large flux of neutrinos from the galactic center, dominating over signals in more standard detection channels. 
Fluxes of the more readily observed $e^\pm$, antiprotons, and photons are  subleading.  
Observational bounds on the flux of cosmic antiprotons are 
stringent, and despite the relatively small production rate of antiprotons in
this class of model, these observations can become constraining if the 
lifetime of the dark matter particle is of order $10^{25}$ seconds.  However, 
the large astrophysical uncertainties in the galactic propagation model for 
antiprotons make it difficult to draw any robust conclusions from the 
antiproton spectrum.  
Moreover, such a dark matter particle naturally 
interacts very weakly with nucleons, giving null predictions
for direct detection experiments.  In these circumstances,
neutrino telescopes become the most sensitive probe of dark matter.

In the example model we present here, the dark matter candidate is a fermion 
belonging to a hidden sector which is neutral under all standard model symmetries.  
Such a particle naturally couples to the standard model via
the neutrino portal, which is the lowest-dimensional fermionic gauge 
singlet operator.  The coupling through the neutrino portal enables dark
matter to decay to the standard model.  The existence of lighter states
in the dark sector allows the decay channel with a single standard model
neutrino to dominate the final state.  This decay mode results in a 
monoenergetic flux of cosmic neutrinos that can be probed by neutrino 
telescopes.  While we have studied a specific realization of that scenario, 
where the hidden sector consists of a confining $SU(N)$ gauge group together
with a heavy fermion in the adjoint representation, obviously similar 
phenomenology can be obtained in a wider class of models.

Currently, the coupling via the neutrino portal is most robustly
constrained by Super-Kamiokande, which restricts the
dark matter lifetime to be larger than $10^{25}$ seconds.  The PAMELA
measurement of the antiproton flux yields roughly comparable constraints
due to the antiprotons arising from the subleading dark matter decay 
channels containing gauge and Higgs bosons.  These bounds leave a lot 
of room for discovery at the current generation of neutrino telescopes.  
The good angular resolution of both ANTARES and IceCube/DeepCore ensure
that the number of signal counts from the direction of the galactic center
may be larger than or comparable to the expected background from
atmospheric neutrinos.  In the case of ANTARES, the sensitivity to the
dark matter decay rate exceeds that of Super-Kamiokande by a factor of
several.  Furthermore, IceCube/DeepCore has the additional prospect of
rejecting the vast majority of atmospheric neutrino events by observing
the associated muon (which is impossible at ANTARES).  This implies
increased sensitivity by another order of magnitude with respect to
ANTARES.

Our analysis of the sensitivity of ANTARES and IceCube/DeepCore should be considered only as suggestive of the experiments' ultimate capabilities; the operational
parameters of these experiments are subject to ongoing research by the
collaborations.  It is conceivable that designing search algorithms
targeted for a monochromatic neutrino signal can further increase
the sensitivity to our scenario.

\bigskip

{ \Large \bf Acknowledgments}

\smallskip \smallskip

We are grateful to Jim Braun and Tyce DeYoung for providing us with
many details concerning the IceCube experiment.  We would like to
thank Peter Graham, Dan Hooper, Patrick Meade, Michele Papucci, Josh Ruderman, Matt Strassler, 
Scott Thomas, and Tomer Volansky for helpful
conversations. This work was supported in part by DOE grant DE-FG02-96ER40959.



\end{document}